\documentclass[preprint,tightenlines,aps,prd,showpacs,nofootinbib]{revtex4}
\usepackage{graphicx}
\usepackage{amsmath}
\usepackage{amssymb}
\usepackage{bm}

\begin{document}

\title{\mbox{}\\[10pt]
Operator Product Expansion in the \\
Production and Decay of the $X(3872)$
}
\author{Eric Braaten and Meng Lu}
\affiliation{
Physics Department, Ohio State University,
Columbus, Ohio 43210, USA}

\date{\today}
\begin{abstract}
The $X(3872)$ seems to be a weakly-bound hadronic molecule whose
constituents are two charm mesons. Its binding energy is much
smaller than all the other energy scales in QCD. This separation of
scales can be exploited through factorization formulas for
production and decay rates of the $X$. In a low-energy effective field
theory for the constituents of the $X$, the factorization formulas can
be derived using the operator product expansion. The derivations are
carried out explicitly for the simplest effective theory in which
the constituents interact through a contact interaction that
produces a large scattering length. The long-distance factors in the
operator product expansions for various observables are calculated
nonperturbatively in the interaction strength of the contact
interaction. After renormalization of the coupling constant, all
remaining ultraviolet divergences can be absorbed into the
short-distance factors in the operator product expansions.
\end{abstract}

\pacs{12.38.-t, 12.39.St, 13.20.Gd, 14.40.Gx}


\maketitle


\section{Introduction}

The $X(3872)$ is a narrow resonance near $3872$ MeV discovered by
the Belle collaboration in 2003 through its decay into
$J/\psi \, \pi^+ \pi^-$ \cite{Choi:2003ue}.  Its existence was subsequently
confirmed by the CDF, Babar, and D0 collaborations
\cite{Acosta:2003zx,Abazov:2004kp,Aubert:2004ns}.  Its mass $M_X$ is
extremely close to the threshold for the charm mesons $D^{0}$ and
$\bar D^{*0}$ \cite{Olsen:2004fp}:
\begin{equation}
M_{X} - (M_{D^{0}}+M_{D^{*0}})= +0.7 \pm 1.1  \; {\rm MeV}.
\label{mXdiff}
\end{equation}
The $X(3872)$ is narrower than most of the known charmonium states
\cite{Choi:2003ue}:
\begin{equation}
\Gamma_{X} < 2.3 \; {\rm MeV} \hspace{1cm} (90\% \ {\rm C.L.}).
\label{GammaX}
\end{equation}
The observation of the decay $X \to J/\psi \, \gamma$ implies
that the $X$ has charge conjugation quantum number $C=+$
\cite{Abe:2005ix}. Analyses of the discovery decay mode 
$X \to J/\psi \, \pi^+ \pi^-$, including the angular correlations 
between the $J/\psi$ and the pions and the $\pi^+ \pi^-$ invariant mass
distribution, strongly favor the spin and parity quantum numbers 
$J^P =1^+$ \cite{Abe:2005iy}. These properties are compatible with the
identification of $X$ as a weakly-bound molecule whose constituents
are a superposition of charm meson pairs
\cite{Tornqvist:2003na,Close:2003sg,Pakvasa:2003ea,%
Voloshin:2003nt,Wong:2003xk,Braaten:2003he,Swanson:2003tb,Braaten:2004rw,
Tornqvist:2004qy,Braaten:2004fk,Swanson:2004pp,Voloshin:2004mh,%
Braaten:2004ai,Braaten:2005jj,AlFiky:2005jd,Braaten:2005ai,Suzuki:2005ha}:
\begin{equation}
X = \frac{1}{\sqrt{2}}
\left( D^{*0} \bar D^0 + D^0 \bar D^{*0} \right).
\label{Xsuper}
\end{equation}
If this identification is confirmed,
the $X(3872)$ would be the first unambiguously identified member
of a new class of hadrons: {\it mesonic molecules}
\cite{Bander:1975fb,Voloshin:ap,DeRujula:1976qd,Nussinov:1976fg,%
Tornqvist:1993ng}.

If the $X(3872)$ is a weakly-bound mesonic molecule, it shares an
important feature with the simplest baryonic molecule, the deuteron.
Their binding energies are both small compared to the natural energy
scale associated with the exchange of the lightest meson, the pion.
That natural energy scale is $m_\pi^2 / (2M_{12})$, where $M_{12}$
is the reduced mass of the two constituents. The binding energy
2.2~MeV of the deuteron is small compared to the natural scale of
about 20~MeV. The measurement of the mass of the $X$
in Eq.~(\ref{mXdiff}) implies that its binding energy
$(M_{D^{0}}+M_{D^{*0}})-M_{X}$ is between $-2.4$ MeV and
$1.2$ MeV at the 90\% confidence level. The small width in
Eq.~(\ref{GammaX}) further suggests that the mass of the $X$
must be below the threshold for the charm mesons:
$M_X < M_{D^0} + M_{D^{*0}}$.  Thus the binding energy of the $X$ is
small compared to the natural scale of about 10~MeV.
The deuteron has an S-wave coupling to its constituents,
the proton and the neutron.  The quantum numbers $J^{PC} = 1^{++}$
of the $X$ implies that it also has an S-wave coupling to its
constituents.  The combination of the small binding energy
compared to the natural energy scale and the S-wave coupling to the
constituents implies that the deuteron and the $X(3872)$
have {\it universal} properties that are determined by the
large scattering length $a$ of their constituents
\cite{Braaten:2003he}. The simplest example of a universal result
is a simple formula for the binding energy of the molecule:
$E_X = 1/(2 M_{12} a^2)$.
The universality of few-body systems with a large scattering length
has many applications in atomic, nuclear, and particle physics
\cite{Braaten:2004rn}.
The universal features of the $X(3872)$ were first exploited by
Voloshin to describe its decays into $D^0 \bar D^0 \pi^0$ and
$D^0 \bar D^0 \gamma$, which can proceed through decay of the
constituent $D^{*0}$ or $\bar D^{*0}$ \cite{Voloshin:2003nt}.
Universality has also been applied to the
production process $\Upsilon(4S) \to \pi^+\pi^- +X$ \cite{Braaten:2004rw},
to the production process $B \to K+X$ \cite{Braaten:2004fk,Braaten:2004ai},
to the line shape of the $X$ \cite{Braaten:2005jj}, and
to decays of $X$ into $J/\psi$ and pions \cite{Braaten:2005ai}.

The tiny binding energy of the $X(3872)$ provides a new energy scale
that is much smaller than the other scales in QCD, including the
pion mass $m_\pi$ and the scale $\Lambda_{QCD}$ associated with
nonperturbative effects. In Ref.~\cite{Braaten:2005jj},
this separation of scales was exploited
by using {\it factorization formulas} to
separate certain observables into long-distance factors that
involve only energy scales comparable to the binding energy and
short-distance factors that involve all the higher energy scales of
QCD. The long-distance factors can be calculated
using an effective field theory for the constituents of the $X$ that
describes the lowest energy scale.

In this paper, we show how the factorization formulas can be derived
using the {\it operator product expansion} for the effective field theory
that describes the constituents of the $X$. The effective field
theory that describes the $X(3872)$ is complicated by the spin 1 of
the constituent $D^{*0}$ and by its charge conjugation quantum
number $C=+$, which implies that it is the superposition of
$D^{*0} \bar D^0$ and $D^0 \bar D^{*0}$ given in Eq.~(\ref{Xsuper}).
Another complication is that the
$D^0 \bar D^0 \pi^0$ threshold is only about 8 MeV below the $D^{*0}
\bar D^0$ threshold \cite{Suzuki:2005ha}. We will therefore
illustrate the operator product expansion formalism using a simpler
model in which these complications are absent. The simplest such
model is a scalar meson model in which the constituents
are spin-0 mesons with a contact interaction that gives a
large positive scattering length $a$.  The generalization
to a realistic model with charm mesons is then straightforward.

In Sec.~\ref{sec:eft}, we define the minimal charm meson model
that can describe the $X(3872)$ and the simpler scalar
meson model. In Sec.~\ref{sec:OPE}, we explain how the operator
product expansion can be used to separate scales in short-distance
production and decay rates. In Sec.~\ref{sec:long}, we give exact
nonperturbative results for long-distance observables in the scalar
meson model. In Sec.~\ref{sec:short}, we apply the operator product
expansion to short-distance production and decay rates in the scalar
meson model and we calculate the long-distance factors in the
operator product expansion.  
We show that after renormalization of the coupling constant,
all remaining dependence on the ultraviolet cutoff can be eliminated
by renormalization of Wilson coefficients in the operator product 
expansion.  In Sec.~\ref{sec:large}, we show how the 
factorization formulas can be simplified by expanding in inverse powers 
of the large scattering length.  In Sec.~\ref{sec:charmmeson}, 
we extend the results of Secs.~\ref{sec:long}, \ref{sec:short}, 
and \ref{sec:large} to the charm meson model. 
We summarize our results in Sect.~\ref{sec:sum}. 


\section{Effective field theories}
\label{sec:eft}

In this section, we define the scalar meson model
and the minimal charm meson model.  Both of these effective field theories
have an S-wave bound state.  We introduce interpolating fields
for the S-wave bound states.

\subsection{Scalar Meson Model}

\begin{figure}[t]
\includegraphics[width=4cm]{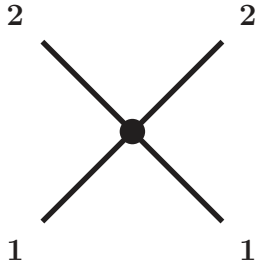}
\caption{ Vertex for the $D_1 D_2$ contact interaction.}
\label{fig:vertex-contact}
\end{figure}

We will illustrate the derivation of factorization formulas
using the operator product expansion in
a simpler model that we call the {\it scalar meson model}.
This model has an S-wave bound state which we will refer to as $X$.
The constituents of the bound state are scalar
mesons $D_1$ and $D_2$ with masses $M_1$ and $M_2$ satisfying
$M_1 < M_2$.  The scalar meson model is a nonrelativistic quantum field
theory with two complex scalar fields $D_1(\vec r,t)$ and $D_2(\vec r,t)$.
The free terms in the lagrangian are
\begin{eqnarray}
{\cal L}_{\rm free} 
=D_1^\dagger\left(i\frac{\partial}{\partial t}
-M_1+\frac{1}{2M_1}\nabla^2\right)D_1
+D_2^\dagger\left(i\frac{\partial}{\partial t}
-M_2+\frac{1}{2M_2}\nabla^2\right)D_2~.
\label{Lfree}
\end{eqnarray}
A superscript $\dagger$ on a field represents its complex conjugate.
The interaction term in the lagrangian for the scalar meson
model is
\begin{equation}
{\cal L}_{\rm int} = - \lambda_0 D_1^\dagger D_2^\dagger D_1 D_2~.
\label{Lint}
\end{equation}
The coupling constant $\lambda_0$ has mass dimension $-2$.
The subscript on $\lambda_0$ emphasizes that it is a bare coupling
constant that depends on the ultraviolet cutoff on the momenta of
the particles in loop diagrams. If the $D_1^\dagger D_2^\dagger D_1
D_2$ interaction is treated nonperturbatively, there is an S-wave
bound state.

It is convenient to introduce concise notations for the reduced
mass of the $D_1$ and $D_2$ and for the sum of their masses:
\begin{subequations}
\begin{eqnarray}
M_{12} &=& \frac{M_1 M_2}{M_1+M_2},
\label{M12}
\\
M_{1+2} &=&  M_1+M_2 .
\label{M1plus2}
\end{eqnarray}
\end{subequations}
If the scalar meson model is an effective field theory for a
more fundamental theory in which the mesons $D_1$ and $D_2$ interact
by the exchange of other mesons, the natural momentum scale
for low-energy processes is the mass $m$ of the lightest meson
that can be exchanged.  If $m \ll M_{12}$,
the natural energy and momentum scales associated with
the exchange of that meson
are $m^2/M_{12}$ and $m$, respectively. We assume that
the binding energy of the molecule $X$ is small compared to the
natural energy scale:
\begin{equation}
M_{1+2} - M_X \ll m^2/M_{12}~.
\label{EX-small}
\end{equation}
The scalar meson model
describes the threshold region where the invariant mass $M$
of $D_1$ and $D_2$ is very close to $M_{1+2}$:
\begin{equation}
|M - M_{1+2}| \ll m^2/M_{12} .
\label{E-small}
\end{equation}
This constraint restricts the possible scattering states to
$D_1 D_2$.

The bare coupling constant $\lambda_0$ of the scalar meson model
must depend on the ultraviolet cutoff $\Lambda$ in such a way that
low-energy observables are independent of $\Lambda$.
There are many alternative renormalization prescriptions that 
can be used to eliminate the explicit
dependence on $\Lambda$ and $\lambda_0$.  One renormalization 
prescription is to eliminate $\lambda_0$ in favor of a renormalized
coupling constant $\lambda$. Another renormalization
prescription is to eliminate $\lambda_0$
in favor of the scattering length $a$ of the two heavy mesons. The
scattering length can be defined in terms of the T-matrix element
for elastic $D_1 D_2$ scattering with zero relative momentum:
\begin{equation}
{\cal T}_{12 \to 12} (p = 0) \equiv - 8 \pi M_{1+2} \, a .
\label{T:p=0}
\end{equation}
This is the T-matrix element for relativistically normalized
particles in the initial and final states.
If either $D_1$ or $D_2$ is an unstable particle,
there is an inelastic scattering channel for $D_1 D_2$.
This implies that the scattering length $a$ has a negative
imaginary part.  The most convenient renormalization prescription
for our purposes is to eliminate $\lambda_0$ in
favor of the energy $E_{\rm pole}$  at which the Green's function
for $D_1 D_2 \to D_1 D_2$ has a pole \cite{Braaten:2005jj}.  That
energy can be expressed in the form
\begin{equation}
E_{\rm pole} =
M_{1+2}  - \gamma^2/(2 M_{12}) ,
\label{Epole}
\end{equation}
where $\gamma$ is the complex binding momentum:
\begin{equation}
\gamma = \gamma_{\rm re} + i \gamma_{\rm im} .
\end{equation}
Unitarity requires $\gamma_{\rm im}$ to be positive.
We assume that $\gamma_{\rm re}$ is also positive, in which case
the energy $E_{\rm pole}$ is the
complex energy of the unstable bound state we denote by $X$.
The real part of $E_{\rm pole}$ defines the pole mass of the molecule:
\begin{equation}
M_X = M_{1+2}
- (\gamma_{\rm re}^2 - \gamma_{\rm im}^2)/(2 M_{12}) .
\end{equation}
The imaginary part of $E_{\rm pole}$ multiplied by $-2$
can be interpreted as the width of the molecule:
\begin{equation}
\Gamma_X = 2 \gamma_{\rm re} \gamma_{\rm im}/M_{12}.
\end{equation}
The magnitude of the complex parameter $\gamma$
is assumed to be small compared to the natural momentum scale:
$|\gamma| \ll m$.
This implies the condition on the binding energy
in Eq.~(\ref{EX-small}).

\subsection{Minimal Charm Meson Model}
\label{sec:MCMM}

The charm mesons $D^{0}$ and $D^{*0}$ with nonrelativistic energies
and momenta can be described by a nonrelativistic quantum field
theory with a complex spin-0 field $D(\vec r,t)$ and a 3-component
complex spin-1 field $\vec D(\vec r,t)$.
Their antiparticles $\bar{D}^0$ and
$\bar{D}^{*0}$ can be described by corresponding fields $\bar D(\vec
r,t)$ and $\vec {\bar D}(\vec r,t)$. The free terms in the
lagrangian density for these particles are
\begin{eqnarray}
\mathcal L_{\text{free}} &=&
D^\dagger\left(i\frac{\partial}{\partial t}
- M_{D^0}+\frac{1}{2M_{D^0}}\nabla^2\right) D
+ \bar{D}^\dagger\left(i\frac{\partial}{\partial t}
-M_{D^0}+\frac{1}{2M_{D^0}}\nabla^2\right)\bar D
\nonumber
\\
&&+{\vec D}^\dagger\cdot\left(i\frac{\partial}{\partial t}
-M_{D^{*0}}+\frac{1}{2M_{D^{*0}}}\nabla^2\right)\vec D
+\vec{\bar D}^\dagger\cdot\left(i\frac{\partial}{\partial t}
-M_{D^{*0}}+\frac{1}{2M_{D^{*0}}}\nabla^2\right)\vec{\bar D}~.
\label{Lfree-Charm}
\end{eqnarray}

The simplest interaction term that can produce an S-wave bound state
in the $C=+$ channel is
\begin{equation}
\mathcal L_{\text{int}} =
-\lambda_0\left(\bar D\vec D + D \vec {\bar D}\right)^\dagger
\cdot \left(\bar D\vec D + D \vec {\bar D}\right).
\label{Lint-Charm}
\end{equation}
We will refer to the effective field theory with
lagrangian given by Eqs.~(\ref{Lfree-Charm}) and (\ref{Lint-Charm})
as the {\it minimal charm meson model}.
If the interaction in Eq.~(\ref{Lint-Charm})
is treated nonperturbatively, there is a S-wave bound state 
with spin 1 that
can be identified with the $X(3872)$. The effects of decays of the
$X$ can be taken into account by allowing the coupling
constant $\lambda_0$ in Eq.~(\ref{Lint-Charm}) to have an imaginary
part. 

An ultraviolet cutoff $\Lambda$ is required to regularize ultraviolet
divergences generated by the interaction term in Eq.~(\ref{Lint-Charm}).
The natural scale for the ultraviolet cutoff is the pion mass
$m_\pi$.  The bare coupling constant $\lambda_0$ 
must depend on $\Lambda$ in such a way that
low-energy observables are independent of the cutoff.
There are many alternative renormalization prescriptions that can
be used to eliminate the explicit dependence on $\Lambda$ and $\lambda_0$.
For example, the complex parameter $\lambda_0$ can be eliminated 
in favor of a renormalized coupling constant $\lambda$
or in favor of the complex scattering length of the charm mesons. 
The most convenient renormalization prescription for our purposes
is to eliminate $\lambda_0$ in favor of the mass and width 
of the $X(3872)$ or equivalently the complex binding momentum $\gamma$. 
An alternative statement of this renormalization prescription 
is that the Green's function for $D^{*0}\bar{D}^0 \it D^{*0}\bar{D}^0$ 
has a pole at the energy $E_{\rm pole}$ given by Eq.~(\ref{Epole}).

\subsection{Interpolating fields for $X$}

In the scalar meson model, the local composite operator $D_1^\dagger
D_2^\dagger(x)$ has a nonzero amplitude to create $X$ from the
vacuum. Thus $D_1 D_2(x)$ can be used as an interpolating field for
$X$. The resulting propagator for $X$ is
\begin{equation}
i \Delta_X(E,P) =
\int d^4x \, e^{i P\cdot x}
\langle \emptyset |  D_1 D_2(x) D_1^\dagger D_2 ^\dagger(0)
    | \emptyset \rangle,
\label{X-prop}
\end{equation}
where $P \cdot x = P^\mu x_\mu$ and $P^\mu =(E, \vec P \,)$ is
the 4-momentum of the $X$.  The propagator is a function of $E$ and
$P=|\vec P|$. The Galilean invariance of the scalar meson model
implies that it depends only on the combination $E - P^2/(2
M_{12})$. Our renormalization prescription implies that this
propagator at $\vec P \,=0$ has a pole in $E$ at the complex energy
$E_{\rm pole}$ given in Eq.~(\ref{Epole}). The behavior of the
propagator near the pole defines a wavefunction normalization
constant $Z_X$:
\begin{equation}
i \Delta_X(E,0) \longrightarrow
\frac{i Z_X}{E - E_{\rm pole} + i \varepsilon} .
\label{DeltaX-pole}
\end{equation}
Because the composite operator $D_1 D_2$ has mass
dimension 3, the propagator $i\Delta_X(E,P)$ has mass dimension 2
and $Z_X$ has mass dimension 3.

T-matrix elements involving $X$ in the final state can be obtained
from connected Green's functions involving the operator $D_1D_2$
by using the LSZ formalism \cite{Peskin-Schroeder}.
The connected Green's function in momentum space
with an external line associated with a $D_1D_2$ operator
is amputated by multiplying by the
inverse propagator for $X$, evaluated on the energy shell
$E= E_{\rm pole}$, and then multiplied by $Z_X^{1/2}$
to obtain the T-matrix element. 
T-matrix elements involving $X$ in the initial state can be obtained
in a similar way from connected Green's functions involving the
operator $D_1^\dagger D_2^\dagger$.

In the charm meson model, $X(3872)$ is identified as a bound state whose
constituents are the $C=+$ superposition of charm mesons
in Eq.~(\ref{Xsuper}).  A convenient
interpolating field for the $X$ is the local composite operator
$D^i \bar D(x) + D \bar D^i(x)$.  The resulting propagator for the
$X$ is
\begin{equation}
i \Delta^{ij}_X(E,P) =
\int d^4x \, e^{i P \cdot x}
\langle \emptyset |
\left( D^i \bar D(x) + D \bar D^i(x) \right)
\left( D^j \bar D(0) + D \bar D^j(0) \right)^\dagger
    | \emptyset \rangle.
\label{X3872-prop}
\end{equation}
If $\vec P \,=0$, the propagator has a pole at $E=E_{\rm
pole}$. Its behavior near the pole defines a normalization factor
$Z_X$:
\begin{equation}
i \Delta^{ij}_X(E,0) \longrightarrow \frac{i Z_X}{E - E_{\rm pole} +
i \varepsilon}\,\delta^{ij}~. \label{DeltaX3872-pole}
\end{equation}

\section{Operator product expansion}
\label{sec:OPE}

\begin{figure}[t]
\includegraphics[width=3.5cm]{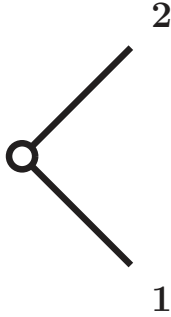}
\caption{ Vertices for the $D_1 D_2$ and $D_1^\dagger D_2^\dagger$
operators. \label{fig:operators}}
\end{figure}

The scalar meson model defined by the lagrangian in Eqs.~(\ref{Lfree})
and (\ref{Lint}) can be a low-energy approximation to a more
fundamental Lorentz-invariant quantum field theory.
If the fundamental quantum field
theory includes a high-energy process that can create $D_1 D_2$ with
invariant mass near $M_{1+2}$, that process will
involve momenta ranging from the highest energy scale
to momenta smaller than $m$. If {\it long-distance}
effects involving momenta much smaller than $m$ can
be separated from {\it short-distance} effects involving momenta of
order $m$ and larger, we can use the scalar meson model to calculate the
long-distance effects. Expressions for physical quantities in
which short-distance effects and long-distance effects are separated
into multiplicative factors are called {\it factorization formulas}.
The tool required to separate long-distance
effects from short-distance effects is the {\it operator product
expansion}.

\subsection{Short-distance production processes}

The general production process for $D_1 D_2$ has the form
$A \to B+D_1 D_2$, where $A$ and $B$ each represent one or more particles.
There can be analogous production processes for $X$. We consider the
production of $D_1 D_2$ near their threshold, so that their
invariant mass $M$ satisfies the inequality in Eq.~(\ref{E-small}).
We also assume that the relative momentum $\vec p \,$ of the $D_1$
and $D_2$ is small compared to $m$. We call $A \to B+D_1 D_2$ a
{\it short-distance production process} if all the particles in $A$
and $B$ have momenta in the $D_1 D_2$ rest frame that are of order
$m$ or larger. This condition implies that the amplitude for
$A \to B+ D_1 D_2$ can be expanded in powers of the small energy
difference $M-M_{1+2}$ divided by $m^2/M_{12}$ and larger energy
scales and in powers of the small relative momentum $\vec p \,$
divided by $m$ and larger momentum scales. The operator product
expansion can be used to express the T-matrix elements in the forms
\begin{subequations}
\begin{eqnarray}
{\cal T}[A \to B + D_1 D_2]  &=&
\sqrt{4 M_1 M_2} \sum_n {\cal C}_A^{B,n}  \,
\langle D_1 D_2 | {\cal O}_n(x=0) | \emptyset \rangle,
\label{T:A12B}
\\
{\cal T}[A \to B+X]  &=&
\sqrt{2 M_X} \sum_n {\cal C}_A^{B,n} \,
\langle X | {\cal O}_n(x=0) | \emptyset \rangle.
\label{T:AXB}
\end{eqnarray}
\label{T:AB}
\end{subequations}
The sums are over local operators
${\cal O}_n$ in the effective field theory.  They can be restricted
to operators with a nonzero matrix element between $\langle D_1 D_2
|$ and the vacuum state $| \emptyset \rangle$. The arguments of the
operator are the origin $x=0$ of space and time. The operator matrix
elements are evaluated in the rest frame of $D_1 D_2$ or $X$.
We use the standard nonrelativistic normalizations for the
states $\langle D_1 D_2 |$ or $\langle X|$ in the operator matrix
elements.  We use the standard relativistic normalizations for the
initial and final states in the T-matrix elements. The
factors of $\sqrt{4 M_1 M_2}$ and $\sqrt{2 M_X}$ account for the
differences between the normalizations of the states in the
operator matrix elements and the T-matrix elements.
The Wilson coefficients ${\cal C}_A^{B,n}$ in Eq.~(\ref{T:AB})
are functions of the 4-momenta and polarization 4-vectors
of the particles in $A$ and $B$ and of the total 4-momentum
$P^\mu$ of a $D_1 D_2$ pair that is produced exactly at threshold with
invariant mass $M_{1+2}$. They also depend on energy scales of order
$m^2/M_{12}$ and higher and on momentum scales of order $m$ and
higher. The only dependence on whether the final state includes
$D_1 D_2$ or $X$ is in the operator matrix elements.
The leading terms in the expansions of the T-matrix elements
in powers of $M - M_{1+2}$ and
$\vec p \,$ are the terms with the lowest dimension operator
$D_1^\dagger D_2^\dagger(0)$. In Feynman diagrams, the local operator
$D_1^\dagger D_2^\dagger$ is represented by an open dot from which a
$D_1$ line and a $D_2$ line emerge, as illustrated in
Fig.~\ref{fig:operators}. The Feynman rule for this vertex is 1.

The operator product expansions in Eqs.~(\ref{T:AB}) provide
the desired separation of long-distance effects and
short-distance effects only if the local operators ${\cal O}_n$
are chosen to be renormalized operators that have
ultraviolet-finite matrix elements.
The simplest local composite operators, such as $D_1 D_2$,
generally have matrix elements
that are ultraviolet divergent.  However the multiplicative
renormalizability of local composite operators implies
that the relation between the simple local operators
${\cal O}_n$ and their renormalized counterparts
${\cal O}^{(R)}_n$ can be expressed as
\begin{equation}
{\cal O}_n(x) =
\sum_m \left( Z^{-1} \right)_{nm} {\cal O}^{(R)}_m(x) ,
\label{oprenorm}
\end{equation}
where $Z$ is an infinite-dimensional matrix of
renormalization constants.
The separation of short-distance effects and long-distance
effects in the T-matrix elements in Eqs.~(\ref{T:AB}) is
accomplished by making the substitutions in Eq.~(\ref{oprenorm})
for the operators ${\cal O}_n$.
The long-distance factors are matrix elements
of the renormalized local operators ${\cal O}^{(R)}_m(x=0)$.
The short-distance factors are the coefficients of these
matrix elements, which are sums of products of Wilson coefficients
${\cal C}_A^{B,n}$ and renormalization constants $(Z^{-1})_{nm}$.
We will find it more convenient to work with simple
composite operators rather than renormalized operators.
The combination of the operator product expansion and the
multiplicative renormalizability of these operators will be used
to separate short-distance effects from long-distance effects.

\subsection{Short-distance decay processes}

The fundamental theory may also allow transitions $D_1 D_2 \to C$
from $D_1 D_2 $ with invariant mass $M_C$
satisfying $|M_C - M_{1+2}| \ll m^2/M_{12}$
to a final state $C$ that includes
particles other than $D_1$ and $D_2$.
If $M_C = M_X$, there can be
analogous transitions $X \to C$. If the sum of the masses of the
particles in $C$ is substantially smaller than $M_{1+2}$, some of
the particles in $C$ must emerge with large momenta. We define a
{\it short-distance transition} to be one for which all the
particles in $C$ have momenta in the center-of-momentum frame
that are of order $m$ or larger. The T-matrix element for such a
process can be expanded in powers of the small energy difference
$M_C - M_{1+2}$ divided by $m^2/M_{12}$ and larger energy scales
and in powers of the small relative momentum $\vec p \,$
of the $D_1$ and $D_2$ divided by $m$ and larger momentum scales.
The operator
product expansion can be used to express the
T-matrix elements in the forms
\begin{subequations}
\begin{eqnarray}
{\cal T}[D_1 D_2 \to C]  &=&
\sqrt{4 M_1 M_2} \sum_n {\cal C}_n^C \,
\langle \emptyset | {\cal O}_n(x=0) | D_1 D_2 \rangle ,
\label{T:12C}
\\
{\cal T}[X \to C]  &=& \sqrt{2 M_X} \sum_n {\cal C}_n^C \,
\langle \emptyset | {\cal O}_n(x=0) | X \rangle . \label{T:XC}
\end{eqnarray}
\label{T:C}
\end{subequations}
The sums over local operators ${\cal O}_n$ of the $D_1D_2$ model can
be restricted to those with a nonzero matrix element between the
vacuum $\langle \emptyset |$ and $| D_1 D_2 \rangle$. The only
dependence on the initial states is in the operator matrix elements.
The leading terms in the expansions of the T-matrix elements
in powers of $M_C - M_{1+2}$ and $\vec p \,$ are the terms
with the lowest dimension operator $D_1D_2(0)$.
The complete separation of short-distance and long-distance
effects is accomplished by using Eq.~(\ref{oprenorm}) to eliminate
the operators ${\cal O}_n$ in favor of renormalized operators.

\subsection{Line shape}

If the fundamental theory includes short-distance processes that
allow the production of $X$ via $A \to B + X$ and the decay of $X$
via $X \to C$, it also allows the process $A \to B + C$, where $C$
represents the same particles but with a variable invariant mass
$M_C$ instead of $M_X$. This process has a resonant enhancement when
$M_C$ is near the $D_1 D_2$ threshold as specified by
Eq.~(\ref{E-small}). If each of the particles in $A$ and $B$ is
well-separated in momentum space from each of the particles in $C$,
the T-matrix element for this process can be described within the
effective field theory by a double operator product expansion:
\begin{eqnarray}
{\cal T}[A \to B + C]  &=& {\cal C}_A^{B,C} +
\sum_{m,n} {\cal C}_A^{B,n} {\cal C}_m^C \int d^4x \, e^{i P \cdot x}
\langle \emptyset | {\cal O}_m(x) {\cal O}_n(0) | \emptyset \rangle.
\label{T:ABCM}
\end{eqnarray}
The sum over operators ${\cal O}_m$ can be restricted to
those with a nonzero matrix element between $\langle D_1 D_2 |$ and
the vacuum state $| \emptyset \rangle$. The sum over operators
${\cal O}_n$ can be restricted to those with a nonzero matrix
element between the vacuum $\langle \emptyset |$ and $| D_1 D_2
\rangle$. The Wilson coefficients ${\cal C}_A^{B,n}$ and ${\cal
C}_m^C$ are the same ones that appear in the operator product
expansions in Eqs.~(\ref{T:AB}) and (\ref{T:C}). In the Fourier
transform of the vacuum--to--vacuum matrix element
in Eq.~(\ref{T:ABCM}), the 4-vector is
$P^\mu = (M_C,\vec 0 \,)$. The leading term in the expansion of the
T-matrix element in powers of $M_C - M_{1+2}$ divided by
$m^2/M_{12}$ and larger energy scales is the term with the lowest
dimension operators $D_1 D_2(x)$ and $D_1^\dagger D_2^\dagger(0)$.
The first term ${\cal C}_A^{B,C}$ on the right side of
Eq.~(\ref{T:ABCM}) takes into account the direct production of $C$
at short distances.  This term can be expanded in powers of the
small energy difference $M_C-M_{1+2}$ divided by $m^2/M_{12}$ and
larger energy scales.  The leading term in the expansion is a
constant independent of $M_C$. The complete separation of
short-distance and long-distance effects is accomplished by using
Eq.~(\ref{oprenorm}) to eliminate the operators ${\cal O}_n$ in
favor of renormalized operators.

\section{Long-distance processes}
\label{sec:long}

In this Section, we give the results for several quantities
in the scalar meson model that depend only on long distances:
the Green's function for $D_1 D_2 \to D_1 D_2$,
the cross section for elastic $D_1 D_2$ scattering,
and the propagator for the bound state $X$.

\subsection{Green's function for $D_1 D_2 \to D_1 D_2$}
\label{sec:Green}

\begin{figure}[t]
\includegraphics[width=12cm]{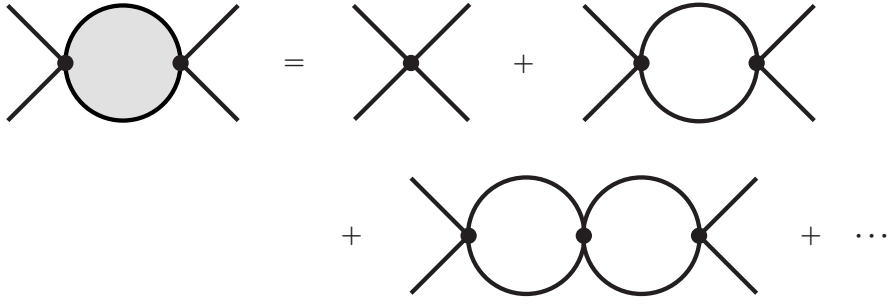}
\caption{
The amputated connected Green's function $i{\cal A}_0(E)$
for $D_1 D_2 \to D_1 D_2$ at $0^{\textrm{th}}$ order in $g$.
It can be obtained by summing
a geometric series of diagrams.
\label{fig:ALO}}
\end{figure}

In the scalar meson model, all the observables for processes
near the $D_1 D_2$ threshold are related in a simple way to the
Green's function for $D_1 D_2 \to D_1 D_2$.
We denote the amputated connected Green's function
for $D_1 D_2 \to D_1 D_2$ by $i {\cal A}_0(E)$,
because it depends only on the total
energy $E$ in the $D_1 D_2$ rest frame.
It can be calculated nonperturbatively by summing the geometric series
represented by Fig.~\ref{fig:ALO} to all orders in $\lambda_0$:
\begin{equation}
i {\cal A}_0(E) = \frac{- i}{1/\lambda_0 -  L_0(E)} ,
\label{A-bare}
\end{equation}
where $i L_0(E)$ is the amplitude for the propagation of the
$D_1 D_2$ pair between successive contact interactions. The function
$L_0(E)$ has an ultraviolet divergence that can be isolated
into an additive term that is independent of $E$.
It can be expressed as
\begin{equation}
L_0(E) = L_0(M_{1+2}) + \frac{M_{12}}{2 \pi}
\sqrt{-2 M_{12} E_{12} - i \varepsilon } ,
\label{L0-sub}
\end{equation}
where $E_{12}$ is the energy relative to the $D_1 D_2$
threshold:
\begin{eqnarray}
E_{12} &=& E - M_{1+2} .
\label{E12}
\end{eqnarray}
The ultraviolet divergence is contained in the term
$L_0(M_{1+2})$. If we use an ultraviolet momentum cutoff $\Lambda
\gg |M_{12} E_{12}|^{1/2}$, this term has a linear ultraviolet
divergence: $L_0(M_{1+2}) = - (M_{12}/\pi^2) \Lambda$.
If we use dimensional regularization, this
term is $L_0(M_{1+2})=0$, because dimensional regularization sets
power ultraviolet divergences equal to 0.

One can identify the combination $1/\lambda_0 - L_0(M_{1+2})$
on the right side of Eq.~(\ref{A-bare})
as a renormalized inverse coupling constant $1/\lambda$.
One possible renormalization prescription is to
eliminate $1/\lambda_0$ in favor of $1/\lambda$.
Another possible renormalization prescription
is to eliminate $\lambda_0$ in favor of the scattering length
$a$ defined in Eq.~(\ref{T:p=0}).  The most convenient 
renormalization prescription for our purposes is to demand
that the pole in the amplitude ${\cal A}_0(E)$ be at the energy 
$E_{\rm pole}$ given in Eq.~(\ref{Epole}).
Inserting the expression for $L_0(E)$ in Eq.~(\ref{L0-sub}) into the
amplitude in Eq.~(\ref{A-bare}) and eliminating $\lambda_0$ in favor
of $\gamma$, the amplitude reduces to
\begin{equation}
{\cal A}_0(E)  =
\frac{2 \pi/M_{12}}{- \gamma + \sqrt{- 2 M_{12} E_{12} - i \varepsilon}} .
\label{A0-ren}
\end{equation}
The renormalized expression for the amplitude ${\cal A}_0(E)$
in Eq.~(\ref{A0-ren}) follows from the renormalization prescription
in Eq.~(\ref{Epole}) and is independent of the regularization scheme.
The expression for the bare coupling constant is
\begin{equation}
\lambda_0 = \frac{1}{L_0(M_{1+2})
+ M_{12}(\gamma_{\rm re} + i \gamma_{\rm im})/(2 \pi)}.
\label{lam0}
\end{equation}
With dimensional regularization, $L_0(M_{1+2}) = 0$,
so Eq.~(\ref{lam0}) gives a finite relation between $\lambda_0$
and the binding momentum $\gamma$: 
$\lambda_0 = 2 \pi/(M_{12} \gamma)$.
We will see later that that the naive use of dimensional
regularization can be misleading.

\subsection{Elastic $D_1 D_2$ Scattering}
\label{sec:D1D2Scattering}

We can use the amplitude in Eq.~(\ref{A0-ren}) to determine the
T-matrix element for the elastic scattering of $D_1$ and $D_2$ with
relative momentum $p$. In the $D_1 D_2$ center-of-momentum frame,
the total energy of the $D_1$ and $D_2$ is
\begin{equation}
E_{\rm cm}(p) = M_{1+2} + p^2/(2 M_{12}) .
\label{Ecm}
\end{equation}
The energy variable $E_{12}$ defined in Eq.~(\ref{E12})
reduces to $E_{12} = p^2/(2 M_{12})$. The T-matrix
element is obtained by evaluating the amplitude ${\cal A}_0(E)$ in
Eq.~(\ref{A0-ren}) at the energy $E_{\rm cm}(p)$ in Eq.~(\ref{Ecm})
and multiplying by the factor $4 M_1 M_2$ to account for the
relativistic normalization of states:
\begin{equation}
{\cal T}_{12 \to 12}(p)  =
- \frac{8 \pi M_{1+2}}{\gamma + i p} .
\label{T0-ren}
\end{equation}
The complex $D_1 D_2$  scattering length defined by
Eq.~(\ref{T:p=0}) is therefore simply
\begin{equation}
a = 1/(\gamma_{\rm re} + i \gamma_{\rm im})  .
\label{dGam:cmm}
\end{equation}

We obtain the cross section for elastic $D_1D_2$ scattering
by squaring the T-matrix element, integrating over the phase space
of the $D_1$ and $D_2$ in the final state, and multiplying by a
flux factor.  The energy $E_{\rm cm}(p)$ in the $D_1 D_2$ rest frame
is assumed to be close to $M_{1+2}$,
as specified by the condition in Eq.~(\ref{E-small}).
The product of the phase space factor
$\lambda^{1/2}(E_{\rm cm}(p),M_1,M_2)/(8 \pi M_{1+2}^2)$
and the flux factor $1/(4 M_{1+2} p)$ can therefore be
approximated by $1/(16 \pi M_{1+2}^2)$.
The cross section is
\begin{equation}
\sigma[D_1D_2(\vec p\,) \to D_1 D_2] =
\frac{4 \pi}{|\gamma_{\rm re} + i(\gamma_{\rm im} + p)|^2} .
\label{sig12}
\end{equation}
The argument $(\vec p\,)$ of $D_1 D_2$ in
the initial state implies that the $D_1$ and $D_2$ have momenta
$-\vec p\,$ and $+\vec p\,$, respectively. The momenta of the $D_1$
and $D_2$ in the final state are not specified because they have
been integrated over.

\subsection{Propagator for $X$}
\label{sec:Xprop0}

If the local composite operator
$D_1 D_2(x)$ is used as an interpolating field for $X$,
the propagator for $X$ is given in Eq.~(\ref{X-prop}).
The diagrams for the propagator of $X$
are shown in Fig.~\ref{fig:Xprop}.
In the rest frame $P=0$, these diagrams form a geometric series
whose sum is
\begin{equation}
i\Delta_X(E,0) = \frac{iL_0(E)}{1 - \lambda_0 L_0(E)}.
\label{DeltaX-0}
\end{equation}
This propagator has a pole in $E$ at the same energy $E_{\rm
pole}$ given in Eq.~(\ref{Epole}) as the amplitude ${\cal A}_0(E)$
in Eq.~(\ref{A-bare}). Near the pole, the behavior of the propagator
at $P=0$ is given in Eq.~(\ref{DeltaX-pole}). Using $L_0(E_{\rm
pole}) = 1/\lambda_0$, we determine the wavefunction normalization
factor to be
\begin{equation}
Z_X = \frac{2 \pi \gamma}{M_{12}^2\lambda_0^2} .
\label{ZX}
\end{equation}

Using the expression for ${\cal A}_0(E)$ in Eq.~(\ref{A-bare}),
the propagator for $X$ in Eq.~(\ref{DeltaX-0}) can be expressed as
\begin{equation}
\Delta_X(E,0) =
-  {\cal A}_0(E) \frac{L_0(E)}{\lambda_0} .
\label{DeltaX}
\end{equation}
The alternative expression for ${\cal A}_0(E)$ in Eq.~(\ref{A0-ren})
shows that, after renormalization of the coupling constant,
it does not depend on the ultraviolet cutoff $\Lambda$.
Thus the propagator in Eq.~(\ref{DeltaX}) depends on $\Lambda$ only
through the factor $L_0(E)/\lambda_0$.  That there is some
dependence on $\Lambda$
is not a surprise, because we have used the simple
composite operator $D_1 D_2(x)$ as the interpolating field for $X$
rather than a renormalized operator.  As we shall see in 
Sec.~\ref{sec:short}, $\lambda_0 D_1 D_2(x)$ is a renormalized operator 
whose matrix elements between the vacuum $\langle \emptyset |$
and $| D_1 D_2 \rangle$ or $| X \rangle$ do not depend 
on the ultraviolet cutoff. 
The propagator for the renormalized operator
$\lambda_0 D_1 D_2(x)$ is obtained by multiplying
the expression in Eq.~(\ref{DeltaX}) by $\lambda_0^2$.
But this propagator depends on
$\Lambda$ through the factor $\lambda_0 L_0(E)$.  To obtain a renormalized
propagator that does not depend on $\Lambda$, one must add
the $\Lambda$-dependent constant $i \lambda_0$
to the propagator for the renormalized operator $\lambda_0 D_1 D_2(x)$:
\begin{equation}
i \lambda_0^2 \, \Delta_X(E,0)  + i \lambda_0 = - i {\cal A}_0(E)~.
\label{DeltaX:ren}
\end{equation}
Thus the renormalized propagator is essentially just the amplitude 
${\cal A}_0(E)$ in Eq.~(\ref{A0-ren}). 
The need for adding the constant term in Eq.~(\ref{DeltaX:ren})
is related to the fact that
if an external source coupled to a composite operator
is added to the lagrangian, renormalization sometimes requires
the addition of terms with higher powers of the source \cite{Zinn-Justin}.
For example, the addition of the term
$J^\dagger D_1 D_2 + {\rm h.c.}$
creates new ultraviolet divergences that can only be
cancelled by a $J^\dagger J$ term.
Such a term is required even in the absence of any interactions.
To implement the LSZ prescription for T-matrix elements 
for processes with $X$ in the initial or final state, it is not 
necessary to use a renormalized propagator.
We will use the unrenormalized propagator for $X$
in Eq.~(\ref{DeltaX}) for this purpose.

\section{Short-distance processes}
\label{sec:short}

In this Section, we consider processes in the scalar meson model
that involve both short-distance and long-distance effects:
short-distance production rates, short-distance decay rates,
and the line shape of the bound state in a short-distance decay mode.
We use the operator product expansion to derive factorization
formulas in which those short-distance effects and long-distance
effects are separated.  After renormalization of the coupling constant, 
all remaining depedence on the ultraviolet cutoff can be removed by
renormalization of the Wilson coefficients in the operator product expansion.

\subsection{Short-distance Production of $X$ and $D_1 D_2$}
\label{sec:Xprod0}

\begin{figure}[t]
\includegraphics[width=8cm]{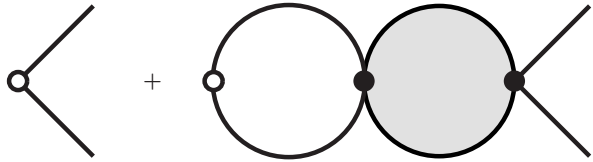}
\caption{
Diagrams at 0$^{\rm th}$ order in $g$ for the vacuum--to--$D_1D_2$
matrix element of the operator $D_1^\dagger D_2^\dagger$.
\label{fig:OLO}}
\end{figure}

We consider the short-distance production processes
$A \to B + D_1 D_2 $ and $A \to B + X$, where $A$ and $B$
both represent one or more particles whose momenta
in the $D_1 D_2$ or $X$ rest frame are all of order $m$ or larger.
The operator product expansions of the
T-matrix element for such short-distance processes are given in
Eqs.~(\ref{T:AB}).
The leading terms in the expansions are the ones
with the local operator $D_1^\dagger D_2^\dagger(0)$. If we keep
only these leading terms, the Lorentz-invariant T-matrix elements
reduce to
\begin{subequations}
\begin{eqnarray}
{\cal T}[A \to B + D_1 D_2]  &=&
\sqrt{4 M_1 M_2} \, {\cal C}_A^{B,12} \,
\langle D_1 D_2 | D_1^\dagger D_2^\dagger(0)
    | \emptyset \rangle,
\label{TADDpB}
\\
{\cal T}[A \to B + X]  &=& \sqrt{2 M_X} \, {\cal C}_A^{B,12} \,
\langle X | D_1^\dagger D_2^\dagger(0) | \emptyset \rangle.
\label{TAXpB}
\end{eqnarray}
\label{TApB}
\end{subequations}

We first consider the T-matrix element for the production of
$D_2 D_2$ via the short-distance process $A \to B + D_1 D_2 $.
We take the $D_1$ and $D_2$ in Eq.~(\ref{TADDpB})
to have relative momentum $\vec p \,$ in the $D_1 D_2$ rest frame
and unspecified total momentum. Their invariant mass is
$M = E_{\rm cm}(p)$, where $E_{\rm cm}(p)$ is given in
Eq.~(\ref{Ecm}). This invariant mass is assumed
to be close to $M_{1+2}$, as specified by the inequality in
Eq.~(\ref{E-small}). The Feynman diagrams for the
vacuum--to--$D_1D_2$ matrix element, which are shown in
Fig.~\ref{fig:OLO}, form a geometric series. The matrix element in
Eq.~(\ref{TADDpB}) is therefore
\begin{equation}
\langle D_1 D_2(\vec p \,) | D_1^\dagger D_2^\dagger(0)
    | \emptyset \rangle
= \frac{1}{1 - \lambda_0 L_0(E_{\rm cm}(p))}.
\label{me-DDdag}
\end{equation}
The argument $(\vec p\,)$ of the $D_1 D_2$ state
implies that the $D_1$ and $D_2$ have momenta $-\vec p\,$
and $+\vec p\,$, respectively.  The sum of the diagrams
in Fig.~\ref{fig:OLO} differs from the geometric series of
diagrams for $i {\cal A}_0(E)$
in Fig.~\ref{fig:ALO} only by the multiplicative
factor $1/(- i \lambda_0)$.  Since ${\cal A}_0(E)$,
which after renormalization of the coupling constant
is given by Eq.~(\ref{A0-ren}),
does not depend on the ultraviolet cutoff,
the operator $\lambda_0 D_1^\dagger D_2^\dagger(x)$
is a renormalized operator whose matrix
elements do not depend on the ultraviolet cutoff. The T-matrix
element in Eq.~(\ref{TADDpB}) can be separated into a short-distance
factor and a long-distance factor by taking the long-distance factor
to be the matrix element of the renormalized operator:
\begin{equation}
{\cal T}[A \to B + D_1D_2(\vec p \,)] =
 - \sqrt{4M_1M_2} \, \big( {\cal C}_A^{B,12}/\lambda_0 \big)  \,
{\cal A}_0(E_{\rm cm}(p)) .
\label{T:A12B-LO}
\end{equation}
After renormalization of the coupling constant,
${\cal A}_0(E_{\rm cm}(p))$ is given by the expression in 
Eq.~(\ref{A0-ren}), which does not depend on the 
ultraviolet cutoff $\Lambda$.  The T-matrix element in 
Eq.~(\ref{T:A12B-LO}) will not depend on $\Lambda$
if the short-distance factor ${\cal C}_A^{B,12}/\lambda_0$
does not depend on $\Lambda$.  Equivalently, the dependence on 
$\Lambda$ can be removed by a multiplicative renormalization of the 
Wilson coefficient ${\cal C}_A^{B,12}$.

\begin{figure}[t]
\includegraphics[width=8cm]{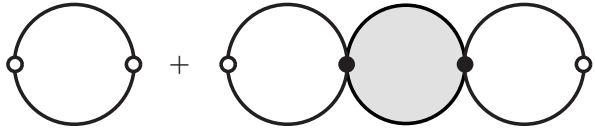}
\caption{
Feynman diagrams for the $X$ propagator at 0$^{\rm th}$ order in $g$.
The interpolating field for the $X$ is $D_1 D_2(x)$.
\label{fig:Xprop}}
\end{figure}

We next consider the T-matrix element for the production of $X$
via the short-distance process $A \to B + X$.
The operator product expansion of the T-matrix element
for this process is given in Eq.~(\ref{TAXpB}).
The vacuum--to--$X$ matrix element can be obtained
via the LSZ formalism from the
connected Green's function for the operator $D_1^\dagger D_2^\dagger$
that acts on the vacuum and an operator $D_1 D_2$ associated
with the $X$ in the final state.
This connected Green's function is identical to the propagator
$ i \Delta_X(E,0)$ given in Eq.~(\ref{DeltaX-0}).
The matrix element in Eq.~(\ref{TAXpB})
is the normalized on-shell amputated connected Green's function.
The Green's function is amputated
by multiplying by the inverse propagator
$[i \Delta_X(E,0)]^{-1}$, which simply gives 1.
In the rest frame of $X$ where its momentum is $\vec P \,=0$,
the Green's function
is put on shell by setting the energy $E$ equal to the energy
$E_{\rm pole}$ in Eq.~(\ref{Epole}), although
the absence of any dependence on $E$ makes this condition moot.
Finally the Green's function is normalized
by multiplying by the factor $Z_X^{1/2}$,
where $Z_X$ is given in Eq.~(\ref{ZX}).
Thus the vacuum--to--$X$ matrix element is simply
\begin{equation}
\langle X | D_1^\dagger D_2^\dagger(0) | \emptyset \rangle =
\frac{(2 \pi \gamma)^{1/2}}{M_{12} \lambda_0} .
\label{XDD0}
\end{equation}
The only factors in the T-matrix element in Eq.~(\ref{TAXpB})
that are sensitive to short distances are the Wilson coefficient
${\cal C}_A^{B,12}$ and the factor of $1/\lambda_0$
from the matrix element.
The T-matrix element  in Eq.~(\ref{TAXpB})
can be expressed as the product
of a short-distance factor and a long-distance factor:
\begin{equation}
{\cal T}[A \to B + X] =
\sqrt{2 M_X} \, \big( {\cal C}_A^{B,12} /\lambda_0 \big) \,
\frac{(2 \pi \gamma)^{1/2}}{M_{12}}.
\label{TAXB}
\end{equation}
The short-distance factor ${\cal C}_A^{B,12}/\lambda_0$
is the same one that appears in Eq.~(\ref{T:A12B-LO}).

The factored expressions for the T-matrix elements in
Eqs.~(\ref{T:A12B-LO}) and (\ref{TAXB}) lead to
factored expressions for the production rates.
The rates for producing $X$ and $D_1 D_2$ are obtained by
squaring the amplitudes and integrating over the appropriate
phase space.  If $A$ consists of a single particle,
the decay rate into $B+X$ can be expressed
in the factored form
\begin{equation}
\Gamma[A \to B + X] =
\Gamma_A^B \, \frac{2 \pi}{M_{12}}
\left| \gamma_{\rm re}^2 + \gamma_{\rm im}^2 \right|^{1/2},
\label{GamX}
\end{equation}
where $\Gamma_A^B$ is a short-distance factor with dimensions
of mass:
\begin{eqnarray}
\Gamma_A^B &=& \frac{M_{1+2}}{M_A M_{12}} \int \frac{d^3P_{1+2}}{(2
\pi)^3 2E_{1+2}} \int \prod_{i \in B} \frac{d^3p_i}{(2 \pi)^3 2E_i}
\left| {\cal C}_A^{B,12}/\lambda_0 \right|^2 (2 \pi)^4
\delta^{(4)}(P_A^\mu - P_{1+2}^\mu 
	- \mbox{$\sum\limits_{i \in B}$}\,p_i^\mu) . 
\nonumber
\\
\label{GamADDB-scalar}
\end{eqnarray}
The 4-momentum $P_{1+2}^\mu$ is that of a $D_1 D_2$ pair
exactly at threshold with invariant mass $M_{1+2}$. We have chosen
the long-distance factor in Eq.~(\ref{GamX}) to be the square of the
long-distance factor in the T-matrix element in Eq.~(\ref{TAXB})
multiplied by $M_{12}$ to make it dimensionless. All factors from
integrating over the phase space of the particles in the final state
are included in the short-distance factor. The differential rate for
producing $D_1 D_2$ with invariant mass $M = E_{\rm cm}(p)$ given by
Eq.~(\ref{Ecm}) can be expressed in the factored form
\begin{equation}
\frac{d \Gamma}{dM}[A \to B + D_1 D_2(\vec p \,)] =
\Gamma_A^B \, \frac{2 p}{|\gamma_{\rm re} + i(\gamma_{\rm im} + p)|^2}.
\label{GamDD}
\end{equation}
The short-distance factor $\Gamma_A^B$ is the same as in
Eq.~(\ref{GamX}). The long-distance factor is the product of 
$4M_1M_2 | {\cal A}_0(E_{\rm cm}(p)) |^2$ from the T-matrix element
in Eq.~(\ref{T:A12B-LO}), the phase space factor 
$\lambda^{1/2}(E_{\rm cm}(p),M_1,M_2)/(8 \pi E_{\rm cm}(p)^2)$, 
the kinematic factor $E_{\rm cm}(p)/\pi$ associated with the 
differential $dM$, and a factor $M_{12}/(2 M_{1+2})$ to compensate 
for the choice of prefactor in Eq.~(\ref{GamADDB-scalar}). 
Using the expression for the invariant mass $E_{\rm cm}(p)$ in
Eq.~(\ref{Ecm}) and replacing $E_{\rm cm}(p)$ 
by $M_{1+2}$ in every factor that is insensitive to $p$, 
the product of the phase space and kinematic
factors can be reduced to $p/(4 \pi^2)$.

\subsection{Short-distance Decay of $X$}
\label{sec:Xdecay0}

The fundamental theory may have short-distance processes
though which $X$ can decay.  We consider the short-distance
decay process $X \to C$, where $C$ represents two or more particles
whose momenta in the $X$ rest frame are of order $m$ or larger.
The operator product expansion for this decay process
is given in Eq.~(\ref{T:XC}).  The leading term
in the expansion is the one with the operator $D_1 D_2(x)$.
If we keep only this term, the T-matrix element reduces to
\begin{equation}
{\cal T}[X \to C]  = \sqrt{2 M_X} \, {\cal C}_{12}^{C} \,
\langle \emptyset | D_1 D_2(0) | X \rangle .
\label{TXCM}
\end{equation}
The $X$--to--vacuum matrix element can be calculated
in a similar way to the vacuum--to--$X$ matrix element
in Eq.~(\ref{XDD0}):%
\footnote{Our notation might suggest that the matrix element in
Eq.~(\ref{0DDX}) is the complex conjugate of the
matrix element in Eq.~(\ref{XDD0}).  However $| X \rangle$
in Eq.~(\ref{0DDX}) is an in state, while $\langle X |$
in Eq.~(\ref{XDD0}) is the hermitian conjugate of an out state.
These two states are related by the S-matrix:
$| X,\ {\rm out} \rangle = {\cal S} | X,\ {\rm in} \rangle$.}
\begin{equation}
\langle \emptyset| D_1 D_2(0) | X \rangle = \frac{(2\pi
\gamma)^{1/2}}
        {M_{12}  \lambda_0} .
\label{0DDX}
\end{equation}
This matrix element depends on the ultraviolet cutoff
only through the factor $1/\lambda_0$.
In the T-matrix element in
Eq.~(\ref{TXCM}), the separation of short-distance
and long-distance effects can be accomplished
by taking the long-distance factor to be the
matrix element of the renormalized operator $\lambda_0 D_1 D_2(0)$:
\begin{equation}
{\cal T}[X \to C]  =
\sqrt{2 M_X} \, \big( {\cal C}_{12}^{C}/\lambda_0 \big) \,
\frac{(2\pi \gamma)^{1/2}}{M_{12}} .
\label{TXCM-fact}
\end{equation}
This T-matrix element will not depend on the
ultraviolet cutoff $\Lambda$ if the short-distance factor
${\cal C}_{12}^{C}/\lambda_0$  does not depend 
on $\Lambda$.  Equivalently, the dependence on $\Lambda$ 
can be removed by a multiplicative renormalization of the 
Wilson coefficient ${\cal C}_{12}^{C}$.

The decay rate of $X$ into the particles represented by $C$ is
obtained by squaring the T-matrix element and integrating over the
phase space of those particles. It can be expressed in the factored
form
\begin{equation}
\Gamma[X \to C]  =
\Gamma^C \; \frac{2\pi}{M_{12}}
\left| \gamma_{\rm re}^2 + \gamma_{\rm im}^2 \right|^{1/2},
\label{GamXCM-fact}
\end{equation}
where $\Gamma^C$ is a short-distance factor with dimensions of
mass:
\begin{equation}
\Gamma_C = \frac{1}{M_{12}} \int \prod_{j \in C} \frac{d^3p_j}{(2
\pi)^3 2E_j} \left| {\cal C}^C_{12}/\lambda_0 \right|^2 (2 \pi)^4
\delta^{(4)} \!( P_{1+2}^\mu - \mbox{$\sum\limits_{j \in C}$}
p_j^\mu) .
\end{equation}
We have chosen the long-distance factor to be the square of
the long-distance factor in the T-matrix element in
Eq.~(\ref{TXCM-fact}) multiplied by $M_{12}$ to make it dimensionless.

The separation of short-distance and long-distance effects
for the transition $D_1(-\vec p \,) D_2(\vec p \,) \to C$
can be accomplished in a similar way.
The operator product expansion for the T-matrix element is
given in Eq.~(\ref{T:12C}).  The T-matrix element can be
expressed as the product of the same short-distance
factor as in Eq.~(\ref{TXCM-fact}) and a long-distance factor
that includes a factor of ${\cal A}_0(E_{\rm cm}(p))$,
where $E_{\rm cm}(p)$ is the energy in Eq.~(\ref{Ecm}).

\subsection{Line shape of $X$ in a Short-distance Decay Mode}
\label{sec:Xlineshape0}

If the fundamental theory includes short-distance processes
that allow the production of $X$ via $A \to B + X$
and the decay of $X$ via $X \to C$, it also allows the process
$A \to B + C$, where $C$ represents the same particles
but with a variable invariant mass $M_C$ instead of $M_X$.
We assume that $M_C$
is near the $D_1 D_2$ threshold as specified by Eq.~(\ref{E-small}).
If every particle in $A$ and $B$ has momentum in the 
rest frame of $C$  of order $m$ or larger,
the T-matrix element for this process
can be expressed as the double
operator product expansion in Eq.~(\ref{T:ABCM}).
The Wilson coefficient $C_A^{B,C}$ can be expanded in powers
of $M_C-M_{1+2}$ divided by $m^2/M_{12}$ and higher energy scales.
The leading term in this expansion is simply a constant
independent of $M_C$.
The leading terms in the double sum come from the
operators ${\cal O}_n(0) = D_1^\dagger D_2^\dagger(0)$
and ${\cal O}_m(x) = D_1D_2(x)$.
According to Eq.~(\ref{X-prop}), the Fourier transform
of the matrix element of these operators in the $C$ rest frame
is just the $X$ propagator, which is given in Eq.~(\ref{DeltaX-0})
or (\ref{DeltaX}), evaluated at $E=M_C$.
Thus the T-matrix element reduces to
\begin{eqnarray}
{\cal T}[A \to B + C]  &=&
{\cal C}_A^{B,C} + {\cal C}_A^{B,12}{\cal C}_{12}^C
\frac{i L_0(M_C)}{1 - \lambda_0 L_0(M_C)} .
\label{T:ABC-bare}
\end{eqnarray}
The Wilson coefficients
and the factor $L_0(E)/\lambda_0$ in the $X$ propagator
depend on the ultraviolet cutoff $\Lambda$. All the dependence on
the energy can be isolated in a term that does not depend on
$\Lambda$ by using the fact that the combination in
Eq.~(\ref{DeltaX:ren}) does not depend on $\Lambda$. By subtracting
and adding $i {\cal C}_A^{B,12} {\cal C}_{12}^C/\lambda_0$ to the
two terms on the right side of Eq.~(\ref{T:ABC-bare}), the T-matrix
element can be expressed as
\begin{eqnarray}
{\cal T}[A \to B + C]  &=& \big( {\cal C}_A^{B,C}
    - i {\cal C}_A^{B,12} {\cal C}_{12}^C/\lambda_0  \big)
- i \big( {\cal C}_A^{B,12} /\lambda_0 \big)
\big( {\cal C}_{12}^C/\lambda_0 \big) \, {\cal A}_0(M_C) .
\label{T:ABC-ren}
\end{eqnarray}
After renormalization of the coupling constant,
${\cal A}_0(E)$ is given in Eq.~(\ref{A0-ren}).
The short-distance factors ${\cal C}_A^{B,12}/\lambda_0$
and ${\cal C}_{12}^C/\lambda_0$ in Eq.~(\ref{T:ABC-ren}) cannot depend 
on $\Lambda$, because otherwise the T-matrix elements in 
Eqs.~(\ref{T:A12B-LO}), (\ref{TAXB}), and
(\ref{TXCM-fact}) would depend on $\Lambda$.  Thus the T-matrix element 
in Eq.~(\ref{T:ABC-ren}) will not depend on $\Lambda$ if the constant term
${\cal C}_A^{B,C} - i {\cal C}_A^{B,12} {\cal C}_{12}^C/\lambda_0$
does not depend on $\Lambda$.  Equivalently, the dependence on 
$\Lambda$ can be removed by an additive 
renormalization of the Wilson coefficient ${\cal C}_A^{B,C}$. 
It is convenient to express the T-matrix element in
Eq.~(\ref{T:ABC-ren}) in the form
\begin{eqnarray}
{\cal T}[A \to B + C]  &=&
- i \big( {\cal C}_A^{B,12} /\lambda_0 \big)
\big( {\cal C}_{12}^C /\lambda_0 \big)
\left[ {\cal A}_0(M_C) - (2 \pi/M_{12}) c_A^{B,C} \right]  ,
\label{T:ABC-simple}
\end{eqnarray}
where $c_A^{B,C}$ is a complex constant with dimension of length
that is completely determined by short-distance factors.
The natural scale for $c_A^{B,C}$ is $1/m$, where $m$ is the mass 
of the lightest meson that can be exchanged between $D_1$ and $D_2$.

The factored expression for the T-matrix element in
Eq.~(\ref{T:ABC-simple}) leads to a factored expression for the rate
for $A \to B+C$. The invariant mass distribution of the particles in
$C$ is obtained by squaring the T-matrix element and integrating
over the momenta of all the particles in the final state. It is
convenient to express the phase space integral in an iterated form
corresponding to the production of the particles in $B$ and an
effective particle of mass $M_C$ followed by the decay of that
effective particle into the particles in $C$.  If $A$  is a single
particle, the decay rate is
\begin{eqnarray}
\Gamma[A \to B + C] &=& \frac{1}{2 M_A} \int \frac{d M_C^2}{2\pi}
\int \frac{d^3P_C}{(2 \pi)^3 2E_C} \int \prod_{i \in B} 
\frac{d^3p_i}{(2 \pi)^3 2E_i} (2 \pi)^4 \delta^{(4)}(P_A^\mu -
P_C^\mu - \mbox{$\sum\limits_{i \in B}$} p_i^\mu) \nonumber
\\
&& \times \int \prod_{j \in C} \! \frac{d^3p_j}{(2 \pi)^3 2E_j} (2
\pi)^4 \delta^{(4)}(P_C^\mu - \mbox{$\sum\limits_{j \in C}$}
p_j^\mu) \left| {\cal T}[A \to B + C] \right|^2  .
\end{eqnarray}
The invariant mass $M_C$ of the particles in $C$
can be replaced by $M_{1+2}$ everywhere except in the long-distance
factor of the T-matrix element.
In that long-distance factor, it can be expressed as
\begin{eqnarray}
M_C = M_{1+2} + p_C^2/(2 M_{12}),
\end{eqnarray}
where $p_C^2$ can be positive or negative.  The variable $p_C$
is pure imaginary if $M_C < M_{1+2}$ and real and positive if
$M_C > M_{1+2}$.  The differential decay rate of the particle $A$
for $M_C$ near $M_{1+2}$ reduces to
\begin{subequations}
\begin{eqnarray}
\frac{d\Gamma}{dM_C}[A \to B+C] &=&
\big( \Gamma_A^B \Gamma^C \big) \, 2 \pi
\left| \frac{1}{(\gamma_{\rm re} - |p_C|) + i \gamma_{\rm im}}
    + c_A^{B,C} \right|^2
\hspace{0.7cm} M_C < M_{1+2},
\\
 &=&
\big( \Gamma_A^B \Gamma^C \big) \, 2 \pi
\left| \frac{1}{\gamma_{\rm re} + i (\gamma_{\rm im} + p_C)} + c_A^{B,C} \right|^2
\hspace{1cm} M_C > M_{1+2}.
\end{eqnarray}
\label{dGamABC}
\end{subequations}
The short-distance factors $\Gamma_A^B$ and $\Gamma^C$ are the
same as in Eqs.~(\ref{GamX}), (\ref{GamDD}), and
(\ref{GamXCM-fact}).  All other short-distance effects are contained
in the complex constant $c_A^{B,C}$.  The invariant mass
distribution in Eq.~(\ref{dGamABC}) is continuous at $M_C=M_{1+2}$.

The separation of scales represented by the renormalized operator product
expansions for the T-matrix elements in Eqs.~(\ref{T:A12B-LO}), 
(\ref{TAXB}), (\ref{TXCM-fact}), and (\ref{T:ABC-ren}) can be obscured by
using dimensional regularization.  In a generic regularization scheme
with ultraviolet cutoff $\Lambda$, the short-distance quantities 
$\lambda_0$, ${\cal C}_A^{B,12}$, ${\cal C}_{12}^C$, and ${\cal C}_A^{B,C}$
are insensitive to $\gamma$.  They depend on $\Lambda$ in such a way that 
the combinations ${\cal C}_A^{B,12}/\lambda_0$, ${\cal C}_{12}^C/\lambda_0$, 
and ${\cal C}_A^{B,C}- i {\cal C}_A^{B,12} {\cal C}_{12}^C/\lambda_0$
do not depend on $\Lambda$.  Dimensional
regularization sets power ultraviolet divergences to zero.  
In particular, it sets $L_0(M_{1+2})=0$, so the expression for 
the bare coupling constant in Eq.~(\ref{lam0}) reduces to 
$\lambda_0 = 2 \pi/(M_{12} \gamma)$.  The Wilson coefficients
${\cal C}_A^{B,12}$, ${\cal C}_{12}^C$, and ${\cal C}_A^{B,C}$
do not depend on the ultraviolet cutoff of dimensional
regularization.  Compatibility with other regularization schemes 
requires however that they depend on $\gamma$ in such a way that the 
combinations ${\cal C}_A^{B,12}/\lambda_0$, ${\cal C}_{12}^C/\lambda_0$, 
and ${\cal C}_A^{B,C}- i {\cal C}_A^{B,12} {\cal C}_{12}^C/\lambda_0$
are insensitive to $\gamma$.  This requires that ${\cal C}_A^{B,12}$
and ${\cal C}_{12}^C$ have multiplicative factors of $\gamma^{-1}$
and that ${\cal C}_A^{B,C}$ have an additive term with a factor 
$\gamma^{-1}$.  Thus the Wilson coefficients in dimensional 
regularization are not short-distance factors.
Their dependence on $\gamma$ 
is similar to the dependence of the Wilson coefficients on $\Lambda$ 
in other regularization schemes.  That dependence cancels in the 
combinations of Wilson coefficients and $\lambda_0$ that appear 
in the renormalized operator product
expansions for the T-matrix elements.

\section{Large scattering length expansion}
\label{sec:large}

Effective field theories can exploit a large separation of 
momentum scales by providing a simpler description of the
lowest momentum scale.  Another  important feature of 
effective field theories is that they provide a systematic 
framework for improving the accuracy of the description 
to any desired order in the ratio of the small
momentum scale and higher momentum scales.  In this Section, 
we discuss how the accuracy of the results for the scalar meson model 
in Secs.~\ref{sec:long} and \ref{sec:short} can be systematically 
improved using an expansion in the large scattering length.
We also explain how that expansion can be exploited 
to simplify some of the results in Sec.~\ref{sec:short}.

In the scalar meson model, the smallest momentum scale 
is the scale $|\gamma|$ associated with the large scattering length.
In a more fundamental theory, there may be many larger momentum 
scales, but the most important at low energies is the mass $m$ 
of the lightest meson that can be exchanged between $D_1$ and $D_2$.
The model defined by the lagrangian in Eqs.~(\ref{Lfree}) 
and (\ref{Lint}) reproduces all effects
that are not suppressed by powers of $|\gamma|/m$.
The model can be systematically improved so that it reproduces 
all corrections to any desired order in $|\gamma|/m$.
We will discuss only the improvements required to reproduce 
corrections through first order in $|\gamma|/m$.

The only improvement in the effective theory that is required to 
decrease the errors to second order in $|\gamma|/m$ is to take 
into account the effective range $r_s$ for S-wave scattering.
This parameter can be defined by the low-momentum expansion 
for the inverse of the T-matrix element:
\begin{eqnarray}
\frac{1}{{\cal T}(p)} &=&
- \frac{1}{8\pi M_{1+2}}
\left( \frac{1}{a} + ip - \frac{1}{2} r_s p^2 + \cdots \right)~.
\label{Tinv}
\end{eqnarray}
If we impose the renormalization condition that the Green's function 
for $D_1 D_2 \longrightarrow D_1 D_2$ has a pole at the energy 
$E_{\rm pole}$ given in Eq.~(\ref{Epole}), one expression that 
will give the correct effective range is
\begin{eqnarray}
{\cal A}(E) &=& 
\frac{-2\pi/M_{12}}{(\gamma + ip)[ 1 - r_s (\gamma-ip)/2 ]}~,
\end{eqnarray}
where $p = i\sqrt{ -2M_{12}E_{12} -i\varepsilon }$.
The corresponding T-matrix element for elastic $D_1 D_2$ scattering 
is then
\begin{equation}
{\cal T}_{12 \to 12}(p)  =
\frac{- 8 \pi M_{1+2}}{(\gamma + i p)[ 1 - r_s (\gamma-ip)/2 ]} .
\label{T0-ren:rs}
\end{equation}

In short-distance observables, there are additional terms 
in the expansions in $|\gamma|/m$ coming from higher dimension 
operators in the operator product expansion.  For each additional 
gradient in the operator, the operator matrix element will have 
an additional factor of order $\gamma$.  The dimensions from 
these additional factors of $\gamma$ must be compensated by
factors of $1/m$ in the short-distance coefficients.
Only operators with a single gradient can give contributions 
that are suppressed by one power of $|\gamma|/m$.
The matrix element of the operator $\nabla^i(D_1 D_2)$ 
between $| D_1 D_2(\vec p \,) \rangle$ or $| X \rangle$ 
and the vacuum $\langle \emptyset |$ vanishes in the 
center-of-momentum frame.  The other independent operator 
with a single gradient is $\nabla^i D_1 D_2 - D_1 \nabla^i D_2$.
The matrix element 
$\langle \emptyset | \nabla^i D_1 D_2 - D_1 \nabla^i D_2| X \rangle$
must vanish because the operator is a vector and there are no
vectors associated with the state $| X \rangle$ in its center-of-mass frame.
The matrix element 
$\langle \emptyset | \nabla^i D_1 D_2 - D_1 \nabla^i D_2
| D_1 D_2(\vec p \,) \rangle$ is nonzero and proportional to $p^i$.
This operator gives a term in the T-matrix element for
$A \to B + D_1D_2(\vec p \,)$ in Eq.~(\ref{T:A12B-LO})
that is linear in the momentum $\vec p \,$ but has a suppression 
factor of $1/m$ in the short-distance coefficient.
Higher dimension operators in the operator product expansion
will contribute to the T-matrix elements for $A \to B + X$ 
in Eq.~(\ref{TAXB}), for $X \to C$ in Eq.~(\ref{TXCM-fact}), 
and for $A \to B + C$ in Eq.~(\ref{T:ABC-simple}) only at second 
and higher orders in $|\gamma|/m$.

The systematic expansion in powers of $|\gamma|/m$
can be used to simplify the leading order results
for short-distance observables.
The T-matrix element for $A \to B + C$ in Eq.~(\ref{T:ABC-simple})
has a resonant term ${\cal A}_0(M_C)$ and a nonresonant 
term $c_A^{B,C}$.  The resonant term ${\cal A}_0(M_C)$ 
includes a factor that is of order $1/|\gamma|$ 
when $p_C$ is of order $|\gamma|$. 
The nonresonant term $c_A^{B,C}$ is completely determined by
short-distance effects, so the natural scale for $c_A^{B,C}$ is $1/m$.
For $p_C$ of order $|\gamma|$,
this amplitude is suppressed by $|\gamma|/m$ 
compared to the resonant term in Eq.~(\ref{T:ABC-simple}).
One can therefore set $c_A^{B,C}=0$ by truncating the expansion 
at leading order in $|\gamma|/m$.  
The invariant mass distribution in Eq.~(\ref{dGamABC})
then reduces to
\begin{subequations}
\begin{eqnarray}
\frac{d\Gamma}{dM_C}[A \to B+C] &=&
\big( \Gamma_A^B \Gamma^C \big) \, 
\frac{2 \pi}{(\gamma_{\rm re} - |p_C|)^2 + \gamma_{\rm im}^2}
\hspace{0.7cm} M_C < M_{1+2},
\\
 &=&
\big( \Gamma_A^B \Gamma^C \big) \, 
\frac{2 \pi}{\gamma_{\rm re}^2 + (\gamma_{\rm im} + p_C)^2} 
\hspace{1cm} M_C > M_{1+2}.
\end{eqnarray}
\label{dGamABC:a}
\end{subequations}
This simple factorization formula was first derived in 
Ref.~\cite{Braaten:2005jj}.
If the nonresonant amplitude $c_A^{B,C}$ 
in Eq.~(\ref{dGamABC}) is included, the systematic expansion 
in powers of $|\gamma|/m$ requires that all other terms 
that are first order in $|\gamma|/m$ also be included.
This requires that the effective field theory be improved 
so that it takes into account the effective range.

The above derivation of the simple factorization formula 
in Eq.~(\ref{dGamABC:a})
is much cleaner than the derivation in Ref.~\cite{Braaten:2005jj}.
In Ref.~\cite{Braaten:2005jj}, the authors used an ultraviolet
momentum cutoff $\Lambda$.  They obtained results that did not 
depend on $\Lambda$ by taking the limit $\Lambda \to \infty$.
The expression for the bare coupling constant $\lambda_0$
in Eq.~(\ref{lam0}) has a term $L_0(M_{1+2})$ in the denominator.
Since $L_0(M_C)$ and $L_0(M_{1+2})$ are both linear in $\Lambda$,
the product $\lambda_0 L_0(M_C)$ approaches $1$
in the limit $\Lambda \to \infty$. 
In this limit, the last factor in the second
term on the right side of Eq.~(\ref{T:ABC-bare}) reduces to 
\begin{eqnarray}
\frac{i L_0(M_C)}{1 - \lambda_0 L_0(M_C)} \longrightarrow
- \frac{i}{\lambda_0^2} \, {\cal A}_0(M_C).
\label{LD:limit}
\end{eqnarray}
The factors of $1/\lambda_0$ can be combined with the 
Wilson coefficients ${\cal C}_A^{B,12}$ and ${\cal C}_{12}^C$
to obtain short-distance factors with finite limits as 
$\Lambda \to \infty$. 
In Ref.~\cite{Braaten:2005jj}, the authors omitted the
${\cal C}_A^{B,C}$ term in Eq.~(\ref{T:ABC-bare}).
This gave the simple factorization formula in Eq.~(\ref{dGamABC:a}). 
In retrospect, omitting the ${\cal C}_A^{B,C}$ term in 
Eq.~(\ref{T:ABC-bare}) can be justified by the observation
that the natural scale for the coefficient $c_A^{B,C}$ 
in Eq.~(\ref{T:ABC-simple}) is $1/m_\pi$.  If $m_\pi$ is identified with the 
ultraviolet cutoff $\Lambda$, then $c_A^{B,C} \to 0$ in the limit 
$\Lambda \to \infty$.  The derivation in Ref.~\cite{Braaten:2005jj} 
blurred the distinction between 
the arbitary unphysical ultraviolet cutoff $\Lambda$, which can be taken 
to $\infty$, and the physical short-distance scale $m_\pi$, which is fixed.
By maintaining the distinction between $\Lambda$ and $m_\pi$,
we were able to separate the renormalization of the operator product 
expansion from the expansion in inverse powers of the large scattering 
length and give a much cleaner derivation of the factorization formula.

\section{Minimal Charm meson model}
\label{sec:charmmeson}

In this section, we generalize the results of Secs.~\ref{sec:long},
\ref{sec:short}, and \ref{sec:large} for the scalar meson model 
to the minimal charm
meson model defined by the lagrangian in Eqs.~(\ref{Lfree-Charm})
and (\ref{Lint-Charm}). The bound state in this model is identified
as the $X(3872)$. For simplicity of notation, we 
denote the masses of $D^0$ and $D^{*0}$ by $M_1=M_{D^0}$ and
$M_2=M_{D^{*0}}$. Thus $M_{12}$ is the reduced mass of $D^0$ and
$D^{*0}$ and $M_{1+2}$ is the sum of their masses.

\subsection{Long-distance processes}

The amplitude $L_0(E)$ in the charm meson model
for the propagation of $D^{*0} \bar D^0$ or
$D^{0} \bar D^{*0}$ between contact interactions is given by the
same expression in Eq.~(\ref{L0-sub}) as in the scalar meson model.
The Green's function for $D^{*0} \bar D^0 \to D^{*0} \bar D^0$ is
diagonal in the vector indices of the spin-1 mesons.
The diagonal entries are
\begin{equation}
i {\cal A}(E) =
\frac{- i}{1/\lambda_0 -  2 L_0(E)} .
\label{A-cmm}
\end{equation}
This differs from the expression for $i{\cal A}_0(E)$ 
in Eq.~(\ref{A-bare}) only in
the factor of 2 multiplying $L_0(E)$, which accounts for the fact
that the particles in each of the loops in Fig.~\ref{fig:ALO} can be
either $D^{*0} \bar D^0$ or $D^{0} \bar D^{*0}$. The Green's
functions for $D^{*0} \bar D^0 \to D^{0} \bar D^{*0}$, $D^{0} \bar
D^{*0} \to D^{*0} \bar D^0$, and $D^{0} \bar D^{*0} \to D^{0} \bar
D^{*0}$ are also given by this same expression. If we use the
renormalization prescription that the Green's function in
Eq.~(\ref{A-cmm}) has a pole in $E$ at the energy $E_{\rm pole}$
given in Eq.~(\ref{Epole}), the expression for the diagonal entries
of the Green's function can be reduced to
\begin{equation}
{\cal A}(E)  =
\frac{\pi/M_{12}}{- \gamma + \sqrt{- 2 M_{12} E_{12} - i \varepsilon}} .
\label{A-ren}
\end{equation}
The complex parameter $\gamma$ determines the
mass and width of a bound state with spin 1 that we identify as the
$X(3872)$.

The Green's function ${\cal A}(E)$ in Eq.~(\ref{A-ren})
differs from ${\cal A}_0(E)$ in Eq.~(\ref{A0-ren}) 
by a factor of $1/2$.  The T-matrix element for 
$D^{*0} \bar D^0 \to D^{*0} \bar D^0$ therefore differs from the 
expression in Eq.~(\ref{T0-ren}) by a factor of 1/2. 
The resulting expression for the cross section for elastic
$D^{*0} \bar D^0$ scattering therefore differs by a factor of $1/4$
from the cross section in Eq.~(\ref{sig12}) for the charm meson model:
\begin{equation}
\sigma[D^{*0} \bar D^0(\vec p\,) \to D^{*0} \bar D^0] =
\frac{\pi}{|\gamma_{\rm re} + i(\gamma_{\rm im} + p)|^2} .
\label{sig12:cmm}
\end{equation}
The argument $(\vec p\,)$ of $D^{*0} \bar D^0$ in the initial
state implies that the $D^{*0}$ and $\bar D^0$ have momenta $+\vec
p\,$ and $-\vec p\,$, respectively. In the final state, the relative
momentum of the $D^{*0}$ and $\bar D^0$ have been integrated over.
The cross section for elastic $D^{0} \bar D^{*0}$ scattering and the
cross sections for $D^{*0} \bar D^0 \to D^{0} \bar D^{*0}$ and
$D^{0} \bar D^{*0} \to D^{*0} \bar D^0$ are also given by the
expression on the right side of Eq.~(\ref{sig12:cmm}).

If the local composite operator $D^i \bar D(x) + D \bar D^i(x)$
is used as the interpolating field for the $X$, the propagator for
$X$ is given in Eq.~(\ref{X3872-prop}).
The Feynman diagrams for the propagator $i\Delta^{ij}_X(E,0)$
in the charm meson model differ from those in Fig.~\ref{fig:Xprop}
for the scalar meson model only in that each loop receives
contributions from two pairs of
particles, $D^{*0} \bar D^0$ and $D^{0} \bar D^{*0}$.
Thus the diagonal entries of the $X$ propagator can be obtained
from the propagator in Eq.~(\ref{DeltaX-0}) by replacing $L_0(E)$ by $2L_0(E)$:
\begin{equation}
i\Delta^{ij}_X(E,0) = %
\frac{i\,2L_0(E)}{1 - 2\lambda_0 L_0(E)} \, \delta^{ij}. %
\label{DeltaX-3872}
\end{equation}
The normalization factor defined by Eq.~(\ref{DeltaX3872-pole}) is
\begin{equation}
Z_X = \frac{ \pi \gamma}{M_{12}^2\lambda_0^2} .
\label{Z-3872}
\end{equation}
This differs by a factor of 2 from the normalization factor $Z_X$
in Eq.~(\ref{ZX}) for the scalar meson model.

The minimal charm meson model is an effective field theory that 
exploits the small ratio between the scale $|\gamma|$ associated 
with a large scattering length and all the larger momentum scales 
of QCD.  At very low energy, the most important of these larger
momentum scales is the mass $m_\pi$ of the pion.  
The minimal charm meson model takes into
account all effects that are not suppressed by powers of 
$|\gamma|/m_\pi$.  The model can be systematically improved so 
that it incorporates corrections to any desired order in
$|\gamma|/m_\pi$.  We will discuss only the improvements 
required to take into account corrections that are first order in
$|\gamma|/m_\pi$.

At first order $|\gamma|/m_\pi$, it is necessary to take into account 
not only the large scattering length $a_+$ in the $C=+$ channel
but also the effective range $r_+$.  It is also necessary to take 
into account the scattering length $a_-$ in the $C=-$ channel. 
These parameters can be defined by low-momentum expansions 
of the T-matrix elements analogous to Eq.~(\ref{Tinv}).  
Since $r_+$ and $a_-$ both have dimensions of length, 
the natural scales for these parameters are order $1/m_\pi$.
If we impose the renormalization condition that the $C=+$ channel
amplitude has a pole in the energy at $E_{\rm pole}$ given in
Eq.~(\ref{Epole}), the Green's functions in the two channels 
can be written as
\begin{subequations}
\begin{eqnarray}
{\cal A}_+(E) &=& 
\frac{-2\pi/M_{12}}{(\gamma + ip)[1-r_+(\gamma-ip)/2]}~,
\label{A+}
\\
{\cal A}_-(E) &=& -\left(2\pi/M_{12}\right)a_-~,
\label{A-}
\end{eqnarray}
\end{subequations} 
where $p = i\sqrt{ -2M_{12}E_{12} -i\varepsilon }$ in Eq.~(\ref{A+}).
The expression for the cross section for elastic 
$D^{*0} \bar D^{0}$ scattering that replaces Eq.~(\ref{sig12:cmm}) is
\begin{equation}
\sigma[D^{*0} \bar D^0(\vec p\,) \to D^{*0} \bar D^0] =
\pi \left| \frac{1}{(\gamma + ip)[1-r_+(\gamma-ip)/2]} + a_- \right|^2~,
\label{sig12:cmmrs}
\end{equation}
The cross section for elastic $D^{0} \bar D^{*0}$ scattering is
given by the same expression.
The cross sections for $D^{*0} \bar D^{0} \to D^{0} \bar D^{*0}$ 
and $D^{0} \bar D^{*0} \to D^{*0} \bar D^{0}$ are
given by the same expression except that $a_-$ is replaced by $-a_-$.

\subsection{Short-distance production of $X$, $D^{*0} \bar D^0$,
and $D^0 \bar D^{*0}$}

The operator product expansion for the charm meson model can be used
to separate the rates for short-distance production and decay
processes into short-distance factors and long-distance factors. We
first consider the short-distance production process $A \to B + X$
and the corresponding production processes for $D^{*0} \bar D^0$ and
$D^0 \bar D^{*0}$. A specific example of such a process is the
discovery production process $B^+ \to K^+ + X$. The leading terms in
the operator product expansions for these processes are those with
the operators $D^{i\dagger} \bar D^\dagger(0)$ and $D^\dagger \bar
D^{i\dagger}(0)$. The expressions for the T-matrix elements
analogous to Eqs.~(\ref{TApB}) are
\begin{subequations}
\begin{eqnarray}
{\cal T}[A \to B + D^{*0} \bar D^0]  &=&
\sqrt{4 M_1 M_2}
\left(  {\cal C}_A^{B,i} \,
\langle D^{*0} \bar D^0 | D^{i\dagger} \bar D^\dagger(0)| \emptyset \rangle
\right.
\nonumber
\\
&& \hspace{2.5cm} \left.
+\, \bar {\cal C}_A^{B,i} \,
\langle D^{*0} \bar D^0 | D^\dagger \bar D^{i\dagger}(0)| \emptyset \rangle
\right),
\\
{\cal T}[A \to B + D^0 \bar D^{*0} ]  &=&
\sqrt{4 M_1 M_2}
\left(  {\cal C}_A^{B,i} \,
\langle D^0 \bar D^{*0} | D^{i\dagger} \bar D ^\dagger(0)| \emptyset \rangle
\right.
\nonumber
\\
&& \hspace{2.5cm} \left.
+\, \bar {\cal C}_A^{B,i} \,
\langle D^0 \bar D^{*0} | D^\dagger \bar D^{i\dagger}(0)| \emptyset \rangle
\right),
\\
{\cal T}[A \to B + X ]  &=& \sqrt{2 M_X}
\left(  {\cal C}_A^{B,i} \,
\langle X | D^{i\dagger} \bar D ^\dagger(0)| \emptyset \rangle
+\, \bar {\cal C}_A^{B,i} \,
\langle X | D^\dagger \bar D^{i\dagger}(0)| \emptyset \rangle
\right) .
\end{eqnarray}
\end{subequations}
The matrix elements between the vacuum and the charm meson
states are
\begin{subequations}
\begin{eqnarray}
\langle D^{*0} \bar D^0(\vec p\,,m) | D^{i\dagger} \bar D^\dagger(0)| \emptyset \rangle
&=& \left( 1 +
\frac{\lambda_0 L_0(E_{\rm cm}(p))}{1 - 2 \lambda_0 L_0(E_{\rm cm}(p))}
\right) \varepsilon^{*i}(m),
\\
\langle D^{*0} \bar D^0(\vec p\,,m) | D^\dagger \bar D^{i\dagger}(0)| \emptyset \rangle
&=&
\frac{\lambda_0 L_0(E_{\rm cm}(p))}{1 - 2 \lambda_0 L_0(E_{\rm cm}(p))}
\, \varepsilon^{*i}(m),
\\
\langle D^0 \bar D^{*0}(\vec p\,,m) | D^{i\dagger} \bar D^\dagger(0)| \emptyset \rangle
&=&
\frac{\lambda_0 L_0(E_{\rm cm}(p))}{1 - 2 \lambda_0 L_0(E_{\rm cm}(p))}
\, \varepsilon^{*i}(m),
\\
\langle D^0 \bar D^{*0}(\vec p\,,m) | D^\dagger \bar D^{i\dagger}(0)| \emptyset \rangle
&=& \left( 1 +
\frac{\lambda_0 L_0(E_{\rm cm}(p))}{1 - 2 \lambda_0 L_0(E_{\rm cm}(p))}
\right) \varepsilon^{*i}(m),
\end{eqnarray}
\label{T-DDX}
\end{subequations}
where $\varepsilon^i(m)$ is the polarization vector for the
spin-1 meson. The arguments $(\vec p\,,m)$ of the
$D^{*0} \bar D^0$ and $D^{0} \bar D^{*0}$ states
imply that the spin 1 and spin 0 mesons have momenta $+\vec p\,$
and $-\vec p\,$, respectively,
and that the spin 1 meson has spin quantum number $m$.
The matrix elements between the vacuum
and the $X$ are
\begin{subequations}
\begin{eqnarray}
\langle X(m) | D^{i\dagger} \bar D^\dagger(0)| \emptyset \rangle
&=& \mbox{$\frac{1}{2}$} Z_X^{1/2} \, \varepsilon^{*i}(m),
\\
\langle X(m) | D^\dagger \bar D^{i\dagger}(0)| \emptyset \rangle
&=& \mbox{$\frac{1}{2}$} Z_X^{1/2} \, \varepsilon^{*i}(m),
\end{eqnarray}
\label{me-3872}
\end{subequations}
where $\varepsilon^i(m)$ is the polarization vector for the $X$ and
the normalization constant $Z_X$ is given in
Eq.~(\ref{Z-3872}).
The factors of $1/2$ in Eqs.~(\ref{me-3872}) come from the fact
that the Green's function for the operators
$D^i \bar D(x) + D \bar D^i(x)$ and $D^{i\dagger} \bar D^\dagger(0)$
or $D^\dagger \bar D^{i\dagger}(0)$ are equal to the propagator for the $X$
given in Eq.~(\ref{DeltaX-3872}) multiplied by $1/2$.
The T-matrix elements in Eqs.~(\ref{T-DDX}) can be expressed
in a form in which short-distance and long-distance effects are
separated:
\begin{subequations}
\begin{eqnarray}
{\cal T}[A \to B + D^{*0}\bar D^0(\vec p\,,m) ]  &=&
\sqrt{4 M_1 M_2}
\left[  - \frac{{\cal C}_A^{B,i} + \bar {\cal C}_A^{B,i}}{2 \lambda_0} \,
    {\cal A}(E_{\rm cm}(p))
+ \frac{{\cal C}_A^{B,i} - \bar {\cal C}_A^{B,i}}{2} \right] \varepsilon^{*i}(m),
\label{facDstarDbar}
\nonumber\\
\\
{\cal T}[A \to B + D^0 \bar D^{*0}(\vec p\,,m) ]  &=&
\sqrt{4 M_1 M_2}
\left[  - \frac{{\cal C}_A^{B,i} + \bar {\cal C}_A^{B,i}}{2 \lambda_0} \,
    {\cal A}(E_{\rm cm}(p))
- \frac{{\cal C}_A^{B,i} - \bar {\cal C}_A^{B,i}}{2} \right] \varepsilon^{*i}(m),
\label{facDDstarbar}
\nonumber\\
\\
{\cal T}[A \to B + X(m) ]  &=& \sqrt{2 M_X} \,
\left[ \frac{{\cal C}_A^{B,i} + \bar {\cal C}_A^{B,i}}{2 \lambda_0} \,
\frac{(\pi \gamma)^{1/2}}{M_{12}} \right] \varepsilon^{*i}(m).
\label{facX-3872}
\end{eqnarray}
\label{facXDD-charm}
\end{subequations}
These T-matrix elements do not depend on the ultraviolet cutoff if
$({\cal C}_A^{B,i} + \bar {\cal C}_A^{B,i})/\lambda_0$ and
${\cal C}_A^{B,i} - \bar {\cal C}_A^{B,i}$ do not depend on $\Lambda$.
Equivalently,  their dependence on $\Lambda$ can be eliminated by 
renormalizations of the Wilson coefficients ${\cal C}_A^{B,i}$ 
and $\bar {\cal C}_A^{B,i}$.

Since the T-matrix element for $A \to B + X$ in Eq.~(\ref{facX-3872})
is the product of a short-distance factor and a long-distance factor 
proportional to $\gamma^{1/2}$, the rate can be expressed as the product 
of a short-distance factor and a long-distance factor 
proportional to $|\gamma|$.  If $A$ consists of a single particle, 
the decay rate is
\begin{equation}
\Gamma [ A \to B + X  ] =
\Gamma^{B}_{A} \frac{2\pi}{M_{12}}
|\gamma_{\text{re}}^2 + \gamma_{\text{im}}^2|^{1/2} ~.
\label{GamABX}
\end{equation}
We have chosen the long-distance factor to be the
same as in Eq.~(\ref{GamX}). 

The expressions for the invariant mass distributions for 
$D^{*0}\bar D^0$ and $D^0 \bar D^{*0}$ that follow from the 
T-matrix elements in Eqs.~(\ref{facDstarDbar}) and (\ref{facDDstarbar})
are much more complicated and depend on the types of particles 
in $B$.  However these expressions simplify if we keep only the 
leading terms in the expansions in $|\gamma|/m$.  For $p$ of order 
$|\gamma|$, the resonant amplitude ${\cal A}(E_{\rm cm}(p))$ in
Eqs.~(\ref{facDstarDbar}) and (\ref{facDDstarbar}) has a factor 
of order $1/|\gamma|$.  The nonresonant terms in
Eqs.~(\ref{facDstarDbar}) and (\ref{facDDstarbar}) involve only 
short-distance factors and are insensitive to the scale 
$|\gamma|$. They are therefore suppressed 
relative to the resonant terms by $|\gamma|/m$. If we take 
$p$ to be of order $|\gamma|$ and keep only 
the leading terms in $|\gamma|/m$, the invariant mass distributions 
reduce to
\begin{subequations}
\begin{eqnarray}
\frac{d\Gamma}{dM}[A \to B + D^{*0}\bar D^0(\vec p\,)] &=&
\Gamma^{B}_{A} \, 
\frac{p}{\gamma_{\rm re}^2 + (\gamma_{\rm im} + p)^2},
\label{GamABDD:1}
\\
\frac{d\Gamma}{dM}[A \to B + D^0 \bar D^{*0}(\vec p\,) ] &=&
\Gamma^{B}_{A} \, 
\frac{p}{\gamma_{\rm re}^2 + (\gamma_{\rm im} + p)^2}.
\label{GamABDD:2}
\end{eqnarray}
\label{GamABDD}
\end{subequations}
We have replaced $E_{\rm cm}(p)$ 
by $M_{1+2}$ everywhere except in the long-distance factor. 
The short-distance factors $\Gamma_{A}^{B}$ are the same as in
Eq.~(\ref{GamABX}).  They cancel out of the ratio between 
Eq.~(\ref{GamABDD:1}) or Eq.~(\ref{GamABDD:2}) and Eq.~(\ref{GamABX}).
This ratio differs from the ratio between Eq.~(\ref{GamDD})
and Eq.~(\ref{GamX}) in the scalar meson model by the probability 
$\frac{1}{2}$ for the $D^{*0}\bar D^0$ and $D^0 \bar D^{*0}$
to be in the $C=+$ channel.  

The factorization formulas in Eqs.~(\ref{GamABX}) and (\ref{GamABDD}) 
were first derived in Refs.~\cite{Braaten:2004fk} and 
\cite{Braaten:2004ai} for the case $\gamma_{\rm im}=0$. 
They were applied to the decays
of $B$ mesons to $K+X$, $K+D^{*0}\bar D^0$, and $K + D^0 \bar D^{*0}$.
The factorization formulas were generalized to the case 
$\gamma_{\rm im}>0$ in Ref.~\cite{Braaten:2005jj}.
The short-distance coefficient $\Gamma^{B}_{A}$ in 
Eqs.~(\ref{GamABX}) can be eliminated in favor of the 
decay rate $\Gamma[A \to B + X]$ using Eq.~(\ref{GamABDD}):
\begin{eqnarray}
\frac{d\Gamma}{dM}[A \to B + D^{*0}\bar D^0(\vec p\,)] &=&
\Gamma[A \to B + X] \, 
\frac{M_{12} p}
    {2 \pi \, |\gamma| \, |\gamma + i p|^2}.
\label{GamABDD:3}
\end{eqnarray}
The coefficient of $\Gamma[A \to B + X]$ agrees with
Ref.~\cite{Braaten:2005jj}.  The corresponding coefficient in  
Refs.~\cite{Braaten:2004fk} and \cite{Braaten:2004ai} 
is larger by a factor of 2.  The origin of this discrepancy 
is an error by a factor of $\sqrt{2}$ 
in the coalescence amplitude ${\cal A}[D^{*0} \bar D^0 \to X]$
in Refs.~\cite{Braaten:2004fk} and \cite{Braaten:2004ai}.
They used the universal prediction for this amplitude 
that was derived in Ref.~\cite{Braaten:2004rw}.  
The coalescence amplitude is determined by the residue 
of the pole in the energy for the amplitude for
$D^{*0} \bar D^0 \to D^{*0} \bar D^0$:
\begin{eqnarray}
{\cal A}[D^{*0} \bar D^0 \to D^{*0} \bar D^0] &=&
\frac{- |{\cal A}[D^{*0} \bar D^0 \to X]|^2}
    {2 M_X [E_{12} + 1/(2 M_{12} a^2)]}
    \qquad {\rm as\ } E_{12} \longrightarrow - 1/(2 M_{12} a^2).
\end{eqnarray}
The universal prediction for this amplitude 
was first derived in Ref.~\cite{Braaten:2004rw} and that result 
was used in Refs.~\cite{Braaten:2004fk} and \cite{Braaten:2004ai}.  
The error in ${\cal A}[D^{*0} \bar D^0 \to X]$ in
Ref.~\cite{Braaten:2004rw} came from an error in 
the amplitude ${\cal A}[D^{*0} \bar D^0 \to D^{*0} \bar D^0]$,
which was larger by a factor of 2 than the correct expression in 
Eq.~(\ref{A-ren}).  The same error in the amplitude 
${\cal A}[D^{*0} \bar D^0 \to D^{*0} \bar D^0]$ appears
in Ref.~\cite{Braaten:2004fk}. 

We can exploit the fact that the minimal charm meson model is
an effective field theory for a more fundamental Lorentz-invariant
field theory, namely the Standard Model. This implies that
${\cal C}_A^{B,i} \varepsilon^{*i}(m)$ and 
$\bar {\cal C}_A^{B,i} \varepsilon^{*i}(m)$ 
must have Lorentz-invariant expressions in terms
of the 4-momenta and polarization 4-vectors of the particles in
$A$ and $B$, the 4-vector $P_{1+2}^\mu$, and the polarization
4-vector $\varepsilon^\mu$ whose 3-vector part reduces to
$\varepsilon^i(m)$ in the frame where $P_{1+2}^\mu = (M_{1+2},\vec 0
\,)$.

We proceed to deduce the constraints of Lorentz invariance
on the  discovery production
process $B^+ \to K^+ + X$ and the corresponding production processes
for $D^{*0}\bar D^0$ and $D^0 \bar D^{*0}$. The Lorentz invariance
of the fundamental theory is a particularly powerful constraint in
this case. The only 4-vectors that the short-distance factors can
depend on are the 4-momenta $P_B^\mu$, $P_K^\mu$, and $P_{1+2}^\mu$,
which satisfy $P_B^\mu = P_K^\mu + P_{1+2}^\mu$, and the polarization
4-vector $\varepsilon^\mu$, which satisfies 
$P_{1+2} \cdot \varepsilon=0$. 
The only independent Lorentz scalar that is linear
in $\varepsilon^*$ is $P_B \cdot \varepsilon^*$. 
Inner products of the 4-momenta can be expressed in
terms of the masses $M_B$, $M_K$, and $M_{1+2}$. Thus 
Lorentz invariance implies that the T-matrix elements in 
(\ref{facXDD-charm}) are determined by two complex
constants $C_+$ and $C_-$ defined by
\begin{subequations}
\begin{eqnarray}
\frac{{\cal C}_{B^+}^{K^+,i} + \bar {\cal C}_{B^+}^{K^+,i}}{2
\lambda_0} \varepsilon^{*i}(m) &=& C_+ P_B \cdot \varepsilon^*(m) ,
\\
\frac{{\cal C}_{B^+}^{K^+,i} - \bar {\cal C}_{B^+}^{K^+,i}}{2}
\varepsilon^{*i}(m) &=& C_- P_B \cdot \varepsilon^*(m) .
\end{eqnarray}
\end{subequations}
The T-matrix elements in Eqs.~(\ref{facXDD-charm}) reduce to
\begin{subequations}
\begin{eqnarray}
{\cal T}[B^+ \to K^+ + D^{*0}\bar D^0(\vec p\,,m) ]  &=&
\sqrt{4 M_1 M_2}
\left[  - C_+ \, {\cal A}(E_{\rm cm}(p)) + C_- \right]
P_B \cdot \varepsilon^*(m),
\\
{\cal T}[B^+ \to K^+ + D^0 \bar D^{*0}(\vec p\,,m) ]  &=&
\sqrt{4 M_1 M_2}
\left[  - C_+ \, {\cal A}(E_{\rm cm}(p)) - C_- \right]
P_B \cdot \varepsilon^*(m),
\\
{\cal T}[B^+ \to K^+ + X(m) ]  &=& \sqrt{2 M_X} \,
C_+ \frac{(\pi \gamma)^{1/2}}{M_{12}}
P_B \cdot \varepsilon^*(m).
\end{eqnarray}
\end{subequations}
The decay rate for $B^+ \to K^+ + X$ can be expressed in the
factored form in Eq.~(\ref{GamABX}) with the short-distance factor
\begin{equation}
\Gamma^{K^+}_{B^+} = 
\frac{\lambda^{3/2}(M_B,M_K,M_{1+2})}{64 \pi M_B^3 M_1 M_2}
|C_+|^2.
\end{equation}
The invariant mass
distributions for the charm mesons in the decays of $B^+$ into $K^+
+ D^{*0}\bar D^0$ and $K^+ + D^{0}\bar D^{*0}$ can be expressed in
factored forms analogous to Eq.~(\ref{GamDD}):
\begin{subequations}
\begin{eqnarray}
\frac{d\Gamma}{dM}[B^+ \to K^+ + D^{*0}\bar D^0(\vec p\,)] &=&
\Gamma^{K^+}_{B^+} \, p
\left| \frac{1}{\gamma_{\rm re} + i(\gamma_{\rm im} +
p)} + c^{K^+}_{B^+} \right|^2,
\label{GamADDB:1}
\\
\frac{d\Gamma}{dM}[B^+ \to K^+ + D^0 \bar D^{*0}(\vec p\,) ] &=&
\Gamma^{K^+}_{B^+} \, p
\left| \frac{1}{\gamma_{\rm re} + i(\gamma_{\rm im} +
p)} - c^{K^+}_{B^+} \right|^2,
\end{eqnarray}
\label{GamADDB-charm}
\end{subequations}
where $c^{K^+}_{B^+}= (M_{12}/2\pi)C_-/C_+ $ is a complex constant.  
The nonresonant amplitude $c^{K^+}_{B^+}$ is required by 
the renormalization of the operator product expansion.
It is completely determined by the short-distance coefficients 
$C_-$ and $C_+$ which are insensitive to the 
small momentum scale $|\gamma|$. The smallest momentum scale 
to which they are sensitive is the pion mass
$m_\pi$.  Since $c_{B^+}^{K^+}$ has dimensions of length, the
natural order of magnitude of $c_{B^+}^{K^+}$ is $1/m_\pi$. 
Thus if $p$ is of order $|\gamma|$, the nonresonant terms 
$\pm \, c_{B^+}^{K^+}$ in Eqs.~(\ref{GamADDB-charm}) are suppressed by a
factor of $|\gamma|/m_\pi$ compared to the resonant terms. 
If we keep only the leading terms in the expansion in 
$|\gamma|/m$, we can set $c_{B^+}^{K^+}=0$.
Thus Eqs.~(\ref{GamADDB-charm}) reduce to Eqs.~(\ref{GamABDD}).  
In a systematic expansion in powers of $|\gamma|/m_\pi$,
the nonresonant amplitudes $\pm\, c_{B^+}^{K^+}$ in
Eqs.~(\ref{GamADDB-charm}) would be retained only if all other
effects of the same order in $|\gamma|/m_{\pi}$ were also included.
The effective field theory should be improved 
to take into account the effective range $r_+$ in the $C=+$ channel 
and the scattering length $a_-$ in the $C=-$ channel.
One should also include terms in the operator product expansion 
with the operators $\nabla^i D^j \bar D - D^j \nabla^i \bar D$ and  
$\nabla^i D \bar D^j - D \nabla^i \bar D^j$.  These terms will have 
operator matrix elements with factors of $p^i \epsilon^j$ 
and Wilson coefficients suppressed by $1/m_\pi$. 

We now apply the factorization formulas to the production process
$B^+ \to K^{*+} + X$  and the corresponding production processes for
$D^{*0}\bar D^0$ and $D^0 \bar D^{*0}$. The only 4-vectors that the
short-distance factors can depend on are the 4-momenta $P_B^\mu$,
$P_{K^*}^\mu$, and $P_{1+2}^\mu$ and the polarization 4-vector
$\varepsilon_{K^*}^\mu$ of the $K^*$. Inner products of the
4-momenta can be expressed in terms of the masses $M_B$, $M_{K^*}$,
and $M_{1+2}$. The independent Lorentz scalars that are linear in 
in $\varepsilon_K^{*\mu} \varepsilon^\nu$ are 
$(P_B \cdot \varepsilon_K^*)(P_B \cdot \varepsilon^*)$,
$M_B^2 \, (\varepsilon_{K^*}^* \cdot \varepsilon^*)$, and
$\epsilon_{\mu \nu \alpha \beta} P_B^\alpha P_{1+2}^\beta
	\varepsilon_K^{*\mu} \varepsilon^\nu$. 
Thus the constraint of Lorentz invariance reduces the T-matrix elements 
in Eq.~(\ref{facXDD-charm}) to six complex
constants $D_\pm$, $E_\pm$, and $F_\pm$ defined by
\begin{subequations}
\begin{eqnarray}
\frac{{\cal C}_{B^+}^{K^{*+},i} + \bar {\cal C}_{B^+}^{K^{*+},i}}{2
\lambda_0} \, \varepsilon^{*i}(m) &=& 
\left( D_+ M_B^2 g_{\mu \nu} + E_+  P_{B\mu} P_{B \nu}
+ F_+ \epsilon_{\mu \nu \alpha \beta} P_B^\alpha P_{1+2}^\beta \right)
\varepsilon_K^{*\mu} \varepsilon^{*\nu}(m),
\\
\frac{{\cal C}_{B^+}^{K^{*+},i} - \bar {\cal C}_{B^+}^{K^{*+},i}}{2}
\, \varepsilon^{*i}(m) &=& 
\left( D_- M_B^2 g_{\mu \nu} + E_- P_{B\mu} P_{B \nu} 
+ F_- \epsilon_{\mu \nu \alpha \beta} P_B^\alpha P_{1+2}^\beta \right)
\varepsilon_K^{*\mu} \varepsilon^{*\nu}(m).
\end{eqnarray}
\end{subequations}
The decay rate for $B^+ \to K^{*+} + X$ can be expressed in 
the factored form in Eq.~(\ref{GamABX})
with short-distance factor $\Gamma^{K^{*+}}_{B^+}$.
The invariant mass
distributions for the charm mesons in the decays of $B^+$ into
$K^{*+} + D^{*0}\bar D^0$ and $K^{*+} + D^{0}\bar D^{*0}$ can be
expressed in factored forms analogous to Eqs.~(\ref{GamADDB-charm})
but considerably more complicated. 
If we keep only the leading terms in the expansions in $|\gamma|/m$, 
these expressions reduce to Eqs.~(\ref{GamABDD}).

\subsection{Short-distance decay of $X$}

We now consider the decay of $X$ into a short-distance
decay mode $C$. Examples of such decay modes are
the discovery mode $J/\psi \,\pi^+\pi^-$ and $J/\psi \,\gamma$.
The expression for the
T-matrix element analogous to Eq.~(\ref{TXCM}) is
\begin{eqnarray}
{\cal T}[X \to C]  &=& \sqrt{2 M_X}
\left(  {\cal C}^{C,i} \,
\langle \emptyset | D^{i} \bar D(0)| X \rangle
+\, \bar {\cal C}^{C,i} \,
\langle \emptyset | D \bar D^{i}(0)| X \rangle
\right) .
\end{eqnarray}
The operator matrix elements are given by expressions analogous
to those on the right sides of Eqs.~(\ref{me-3872})
except that $\varepsilon^{*i}(m)$ is replaced by $\varepsilon^i(m)$.
The factored expression for the T-matrix element is
\begin{eqnarray}
{\cal T}[X(m) \to C]  &=& \sqrt{2M_X} \,
\frac{{\cal C}^{C,i} + \bar{\cal C}^{C,i}}{2\lambda_0}
\frac{(\pi \gamma)^{1/2}}{M_{12}} \, \varepsilon^i(m).
\label{TXC}
\end{eqnarray}
The T-matrix element does not depend on the ultraviolet cutoff if
$({\cal C}^{C,i} + \bar{\cal C}^{C,i})/\lambda_0$
does not depend on $\Lambda$.
If we were to consider the process $D^{*0} \bar D^0 \to C$,
we would find that ${\cal C}^{C,i} - \bar{\cal C}^{C,i}$
must also be independent of $\Lambda$. Thus the ultraviolet divergences 
can be removed by renormalizations of the Wilson coefficients 
${\cal C}^{C,i}$ and $\bar{\cal C}^{C,i}$. 
The decay rate for $X \to C$ can be expressed in a factored form
analogous to Eq.~(\ref{GamXCM-fact}):
\begin{equation}
\Gamma[X \to C] =
\Gamma^C \; \frac{2\pi}{M_{12}}
\left| \gamma_{\rm re}^2 + \gamma_{\rm im}^2 \right|^{1/2},
\label{GamXC}
\end{equation}
where $\Gamma^C$ is a short-distance factor with dimension
of mass.  We have chosen the long-distance factor to be the
same as in Eq.~(\ref{GamXCM-fact}). 
The factorization formula in Eq.~(\ref{GamXC}) 
was first derived in Ref.~\cite{Braaten:2005jj}. 

In Ref.~\cite{Braaten:2005ai}, the decay rates of $X(3872)$
into $J/\psi \, \pi^+ \pi^-$, $J/\psi \, \pi^+ \pi^- \pi^0$, 
$J/\psi \, \gamma$, and $J/\psi \, \pi^0 \gamma$
were calculated under the assumption that the decays are 
dominated by a direct coupling of $X$ to $J/\psi$ and the 
vector mesons $\rho$ and $\omega$ followed by the decay 
of the virtual vector mesons into pions and photons.
The decay rates were calculated in terms of 
coupling constants $G_{X\psi\rho}$ and $G_{X\psi\omega}$
and other parameters that were determined by vector meson decays.  
The long-distance scale 
$|\gamma|$ enters the decay rates only through a factor
of $\gamma^{1/2}$ in the coupling constants $G_{X\psi\rho}$ 
and $G_{X\psi\omega}$.  Thus the decay rates in 
Ref.~\cite{Braaten:2005ai} satisfy the factorization formula 
in Eq.~(\ref{GamXC}). 

The Lorentz invariance of the more fundamental theory 
provides constraints on the T-matrix elements for specific
short-distance decay processes. For the decay $X \to J/\psi\gamma$,
the only 4-vectors the short-distance factors can depend on are the
4-momenta $P_{1+2}^\mu$ and $P_\psi^\mu$ or $P_\gamma^\mu$ and the
polarization 4-vectors $\varepsilon_{\psi}^\mu$ and
$\varepsilon_{\gamma}^\mu$ of the $J/\psi$ and the photon. 
The coefficient of 
$\varepsilon_{\psi}^{*\mu} \varepsilon_{\gamma}^{*\nu} 
	\varepsilon^\sigma$ must be a 3-index Lorentz tensor.
There are 6 independent tensors that can be 
constructed from the 4-momenta, the metric, and the Levi-Civita tensor.
Thus Lorentz invariance constrains the T-matrix element to be a linear 
combination of these 6 terms with constant coefficients.
In the model of Ref.~\cite{Braaten:2005ai}, the assumptions 
of the direct coupling of $X$ to $J/\psi \, \rho$ and  $J/\psi \, \omega$
and the vector-meson dominance of the coupling of the photon to hadrons 
were used to reduce the T-matrix to a single term proportional to 
$\epsilon_{\alpha \mu\nu\rho} P_\gamma^\alpha 
\epsilon_{\psi}^{*\mu} \epsilon_{\gamma}^{*\nu}\epsilon^\sigma$. 
For the process $X \to J/\psi \, \pi^+\pi^-$,  the only 4-vectors the
short-distance factors can depend on are 
the 4-momenta $P^\mu_{1+2}$, $P^\mu_{\psi}$, and 
$Q_\pi^\mu = P_{\pi^+}^\mu - P_{\pi^-}^\mu$ and the
polarization 4-vector $\varepsilon_{\psi}^\mu$ of the $J/\psi$.
The coefficient of $\varepsilon_{\psi}^{*\mu} \varepsilon^\nu$ 
must be a 2-index Lorentz tensor.
There are 8 independent tensors that can be
constructed from the 4-momenta, the metric, and the Levi-Civita tensor.
Thus Lorentz invariance constrains the T-matrix element to be a linear 
combination of these 8 terms
with coefficients that are functions of the two
independent Lorentz scalars $Q_\pi \cdot P_\psi$ and 
$P_{1+2} \cdot P_\psi$.
In the model of Ref.~\cite{Braaten:2005ai}, the assumption 
of a direct coupling of the $X$ to $J/\psi \, \rho$ was used 
to reduce the T-matrix element to a single term proportional to 
$\epsilon_{\alpha \beta \mu \nu} (P_{1+2}-P_\psi)^\alpha Q_\pi^\beta
\varepsilon_{\psi}^{*\mu} \varepsilon^\nu$.

\subsection{Line shape of $X$ in a short-distance decay mode}

We now consider the line shape of $X$ in the process $A \to B + C$,
where $C$ is a short-distance decay mode of $X$. An example of such a
process is the discovery process for the $X$:
$B^+ \to K^+ + J/\psi \, \pi^+\pi^-$.
The expression for the T-matrix element analogous to
Eq.~(\ref{T:ABC-bare}) is
\begin{eqnarray}
{\cal T}[A \to B + C]  &=&
{\cal C}_A^{B,C} +
\big( {\cal C}_A^{B,i} {\cal C}^{C,i} +
       \bar{\cal C}_A^{B,i} \bar{\cal C}^{C,i}\big)
iL_0(M_C)
\nonumber
\\
&&+\, \big({\cal C}_A^{B,i} + \bar{\cal C}_A^{B,i}\big)\,
\big({\cal C}^{C,i} + \bar{\cal C}^{C,i}\big)
\frac{i \lambda_0L_0(M_C)^2}
{1-2\lambda_0 L_0(M_C)}~.
\end{eqnarray}
The factorized expression for the T-matrix element
analogous to Eq.~(\ref{T:ABC-ren}) is
\begin{eqnarray}
{\cal T}[A \to B + C]  &=&
-i \,
\frac{ {\cal C}_A^{B,i} + \bar{\cal C}_A^{B,i} } { 2\lambda_0 } \,
\frac{ {\cal C}^{C,i}   + \bar{\cal C}^{C,i}   } { 2\lambda_0 }
{\cal A}(M_C)
+ \big( {\cal C}_A^{B,C} + \Delta {\cal C}_A^{B,C} \big)
\nonumber
\\
&& + \frac{i}{2} \big({\cal C}_A^{B,i} - \bar{\cal C}_A^{B,i}\big) \,
\big({\cal C}^{C,i} - \bar{\cal C}^{C,i}\big) \,
 [ L_0(M_C) - L_0(M_{1+2}) ] ,
\label{TABC:charm}
\end{eqnarray}
where $\Delta {\cal C}_A^{B,C}$ is
\begin{eqnarray}
\Delta {\cal C}_A^{B,C} = &-&i\lambda_0 \,
\frac{{\cal C}_A^{B,i}+\bar{\cal C}_A^{B,i}}{2\lambda_0}
\frac{{\cal C}^{C,i}+\bar{\cal C}^{C,i}}{2\lambda_0}
+\frac{i}{2}
\left({\cal C}_A^{B,i}-\bar{\cal C}_A^{B,i}\right)
\left({\cal C}^{C,i}-\bar{\cal C}^{C,i}\right)
L_0(M_{1+2}).
\end{eqnarray}
After renormalization of the coupling constant,
${\cal A}(M_C)$ is given by the expression in (\ref{A-ren}),
which does not depend on the ultraviolet cutoff $\Lambda$.
The term $L_0(M_C) - L_0(M_{1+2})$, which is given in Eq.~(\ref{L0-sub}),
also does not depend on $\Lambda$.
The combinations $({\cal C}_A^{B,i} + \bar{\cal C}_A^{B,i})\lambda_0$,
$({\cal C}^{C,i}   + \bar{\cal C}^{C,i})/\lambda_0$, 
${\cal C}_A^{B,i} - \bar{\cal C}_A^{B,i}$, and
${\cal C}^{C,i} - \bar{\cal C}^{C,i}$
cannot depend on $\Lambda$, because they appear 
as short-distance factors in other T-matrix 
elements such as those in Eqs.~(\ref{facXDD-charm}) and (\ref{TXC}).
Thus the T-matrix element in Eq.~(\ref{TABC:charm})
will not depend on $\Lambda$ if 
${\cal C}_A^{B,C} + \Delta {\cal C}_A^{B,C}$ does not depend on 
$\Lambda$.  Equivalently, the dependence on $\Lambda$ 
can be removed by an additive renormalization 
of the Wilson coefficient ${\cal C}_A^{B,C}$.

The expression for the rate that follows from the T-matrix element
in Eq.~(\ref{TABC:charm}) is very complicated and depends on the types of
particles in $B$.  The expression simplifies 
if we keep only the leading term in the expansion in $|\gamma|/m_\pi$.
If $|p_C|$ is of order $|\gamma|$, the resonant amplitude 
${\cal A}(M_C)$ has a factor of order $1/|\gamma|$
while the term $L_0(M_C) - L_0(M_{1+2})$ has a factor of order $|\gamma|$.
Additional factors of $|\gamma|$ must be accompanied by 
additional factors of $m_\pi$ in the short-distance  factors.
Thus the second and third terms on the right side of Eq.~(\ref{TABC:charm})
are suppressed by one and two powers of $|\gamma|/m_\pi$, respectively.
If we keep only the leading term in $|\gamma|/m_\pi$,
the invariant mass distribution for $C$ in
the decay of a single particle $A$ into $B + C$ can be expressed 
in the simple factored form
\begin{subequations}
\begin{eqnarray}
\frac{d\Gamma}{dM_C}\,[ A \to B + C] &=&
\Gamma_{A}^{B,C} \, 
\frac{2 \pi}{(\gamma_{\rm re} - |p_C|)^2 + \gamma_{\rm im}^2}
\hspace{0.7cm} M_C < M_{1+2},
\\
 &=&
\Gamma_{A}^{B,C} \, 
\frac{2 \pi}{\gamma_{\rm re}^2 + (\gamma_{\rm im} + p_C)^2}
\hspace{1cm} M_C > M_{1+2}.
\end{eqnarray}
\label{GamABC}
\end{subequations}
The invariant mass distribution is continuous at $M_C
= M_{1+2} = M_{D^0} + M_{D^{*0}}$. 
The factorization formula in Eq.~(\ref{GamABC}) 
was first derived in Ref.~\cite{Braaten:2005jj}.
In contrast to the corresponding factorization formula 
in the scalar meson model which is given in
Eq.~(\ref{dGamABC:a}), the short-distance factor
$\Gamma_{A}^{B,C}$ is not simply the product of the short-distance 
factors $\Gamma_{A}^{B}$ and $\Gamma^C$ in
Eqs.~(\ref{GamXC}), (\ref{GamABDD}), and (\ref{GamXC}).
The reason for this is that
the short-distance factors associated with the initial and final 
states in the T-matrix element in (\ref{TABC:charm}) are connected by 
the vector index $i$.

\section{Summary}
\label{sec:sum}

The $X(3872)$ seems to be a hadronic molecule consisting of 
a $C=+$ superposition of $D^{*0}\bar D^0$ and $D^0 \bar D^{*0}$ 
that are weakly bound in the S-wave channel.
The binding energy and the width of the $X$ can be conveniently 
expressed in terms of the complex binding momentum $\gamma$
defined in Eq.~(\ref{Epole}).  The smallness of $\gamma$ 
compared to the natural scale $m_\pi$ together with the S-wave 
nature of the bound state imply that the $X$ has universal properties
that are completely determined by $\gamma$.  The separation of
scales between $|\gamma|$ and $m_\pi$ can be exploited through 
factorization formulas for the production and short-distance 
decay rates of $X$.  The factorization formulas express 
these rates as the sum of products of short-distance factors 
that are insensitive to $\gamma$ and long-distance factors 
that are completely determined by $\gamma$.

We have shown how the factorization formulas can be derived 
using the operator product expansion for a low-energy 
effective field theory for the charm mesons,
such as the minimal charm meson model. 
Using the operator product expansion, the rates are expressed as sums of
products of Wilson coefficients and matrix elements of operators
in the effective field theory.  In the minimal charm meson model,
the matrix elements can be calculated nonperturbatively
and they depend on the ultraviolet cutoff $\Lambda$.
Some of the dependence on $\Lambda$ can be removed by the renormalization 
of the coupling constant.  This can be accomplished conveniently 
by eliminating the bare coupling constant in favor of $\gamma$.
The remaining dependence on $\Lambda$ can be removed by 
renormalization of the Wilson coefficients in the operator 
product expansion.  After eliminating all dependence on $\Lambda$,
the rate can be expanded in powers of $|\gamma|/m_\pi$.
The leading terms in the expansions are very simple.
The leading terms in the rates for the production processes $A \to B+X$,
$A \to B+D^{*0}\bar D^0$, and $A \to B+D^0 \bar D^{*0}$
are given in Eqs.~(\ref{GamABX}) and (\ref{GamABDD}).
The leading term in the rate for the short distance decay  
process $X \to C$ is given in Eq.~(\ref{GamXC}).
The leading term for the line shape of $X$ in the short-distance 
decay mode $C$ is given in Eq.~(\ref{GamABC}).

Our derivation of the factorization formulas using the operator product
expansion makes it clear how these leading order results can be 
extended systematically to higher orders in $|\gamma|/m_\pi$.
This requires improving the effective field theory and including 
higher dimension operators in the operator product expansion.
If accuracy to $n^{\rm th}$ order in $|\gamma|/m_\pi$ is desired,
the effective field theory must describe the scattering of 
charm mesons to that accuracy and the operator product expansion
must include operators with up to $n$ gradients.
As the order in $|\gamma|/m_\pi$ increases, there are an increasing
number of parameters in the effective field theory and an increasing
number of Wilson coefficients in the relevant terms of the 
operator product expansion.  The difficulty of determining 
all these parameters phenomenologically may limit the utility 
of the expansion to low orders in $\gamma/m_\pi$.

Our derivation of the simple factorization formulas in 
Eqs.~(\ref{GamABX}), (\ref{GamABDD}), (\ref{GamXC}), and (\ref{GamABC}) 
is conceptually cleaner than the previous derivations in 
Refs.~\cite{Braaten:2004fk}, \cite{Braaten:2004ai}, and
\cite{Braaten:2005jj}.  Those previous derivations were awkward 
in that they required taking the limit $\Lambda \to \infty$
while also exploiting the fact that the natural scale 
of the ultraviolet cutoff is $m_\pi$.  In the present derivation,
all dependence on $\Lambda$ is removed analytically through 
renormalization of the coupling constants and through
renormalization of the Wilson coefficients in the operator product
expansion without taking the limit $\Lambda \to \infty$.  
In a subsequent conceptually independent step, 
the rates are expanded in powers of $|\gamma|/m_\pi$.
The leading terms in this expansion give the simple 
factorization formulas.

\begin{acknowledgments}
This research was supported in part by the Department of Energy
under grant DE-FG02-91-ER4069. EB thanks R.~Furnstahl for valuable
discussions.
\end{acknowledgments}


\end{document}